\documentclass[aps,prd,nofootinbib,showpacs,superscriptaddress,twocolumn,floatfix]{revtex4-1}

\usepackage{amsmath,amssymb}
\usepackage[utf8]{inputenc}
\usepackage{graphicx}
\usepackage{mathrsfs}
\usepackage{bm}
\usepackage{bbm}
\usepackage{indentfirst}
\usepackage{epstopdf}
\usepackage{color}
\usepackage{xcolor}
\usepackage{amssymb, amsmath,color, hyperref, graphicx}
\usepackage[mathlines]{lineno}

\usepackage{braket}
\usepackage{placeins}
\usepackage{multirow}
\usepackage{slashed}
\usepackage{physics}

\newcommand{\bec}{\begin{center}}
\newcommand{\eec}{\end{center}}
\newcommand{\beq}{\begin{equation}}
\newcommand{\eeq}{\end{equation}}
\newcommand{\bea}{\begin{eqnarray}}
\newcommand{\eea}{\end{eqnarray}}
\newcommand{\nn}{\nonumber}

\newcommand{\hf}{\frac{1}{2}}

\newcommand{\tB}{{\rm B}}
\newcommand{\tQ}{{\rm Q}}
\newcommand{\tS}{{\rm S}}

\newcommand{\hmu}{{\hat{\mu}}}
\newcommand{\hmuQ}{{\hat{\mu}_{\rm Q}}}
\newcommand{\hmuS}{{\hat{\mu}_{\rm S}}}
\newcommand{\hmuB}{{\hat{\mu}_{\rm B}}}
\newcommand{\muQ}{{{\mu}_{\rm Q}}}
\newcommand{\muB}{{{\mu}_{\rm B}}}
\newcommand{\muS}{{{\mu}_{\rm S}}}

\newcommand{\chiBQ}{\chi^{\rm B Q}_{11}}
\newcommand{\chiQS}{\chi^{\tQ \tS}_{11}}
\newcommand{\chiBS}{\chi^{\tB \tS}_{11}}
\newcommand{\chiB}{\chi^{\tB}_2}
\newcommand{\chiQ}{\chi^{\tQ}_2}
\newcommand{\chiS}{\chi^{\tS}_2}

\begin{document}

\title{Leading-Order QCD Equation of State in Strong Magnetic Fields \\ 
at Nonzero Baryon Chemical Potential}

\author{Heng-Tong Ding}
\address{Key Laboratory of Quark and Lepton Physics (MOE) and Institute of
Particle Physics, Central China Normal University, Wuhan 430079, China}
\author{Jin-Biao Gu}
\address{Key Laboratory of Quark and Lepton Physics (MOE) and Institute of
Particle Physics, Central China Normal University, Wuhan 430079, China}
\author{Arpith Kumar}
\address{Key Laboratory of Quark and Lepton Physics (MOE) and Institute of
Particle Physics, Central China Normal University, Wuhan 430079, China}
\author{Sheng-Tai Li}
\affiliation{Key Laboratory of Quark and Lepton Physics (MOE) and Institute of
Particle Physics, Central China Normal University, Wuhan 430079, China}

\date{\today}% 

%------------------------------------------------------------------------------------
%       abstract
%------------------------------------------------------------------------------------
\begin{abstract}    
We present continuum-estimated $(2+1)$-flavor lattice QCD results for the leading-order Taylor expansion coefficients of the equation of state in strong magnetic fields and at nonzero baryon chemical potential. Simulations employ the highly improved staggered quark (HISQ) action with physical pion masses on lattices of temporal extent $N_\tau = 8,\,12$, covering $145 \lesssim T \lesssim 165~\mathrm{MeV}$ and $eB \lesssim 0.8~\mathrm{GeV}^2$, imposing strangeness neutrality with baseline results at electric charge to baryon number ratio $r = 0.4$.  We determine the $T$--$eB$ dependence of $q_1$ and $s_1$ (electric charge and strangeness chemical potential ratios), pressure coefficient $P_2$, baryon number density coefficient $N_1^\tB$, and energy-like coefficients $\Theta_2$ (trace anomaly), $\epsilon_2$ (energy density), and $\sigma_2$ (entropy density). Magnetic fields induce temperature-band crossings for $q_1$ and $P_2$ and non-monotonic structures in the energy-like coefficients, with $\Theta_2$ at strong fields possibly vanishing or turning negative at higher $T$, indicating dominance of the pressure term over the energy contribution. 
We also examine the $r$-dependence, finding that $r=0$ (charge-neutral matter) shows the most muted magnetic-field enhancement of $P_2$ despite larger $|q_1|$, providing a useful reference for neutron-star–like conditions. 
Comparisons with the hadron resonance gas (HRG) model show qualitative agreement at low $T$ and weak $eB$, with clear deviations near the crossover and at strong fields. These results provide useful input for constraining models and effective theories of QCD matter in strong magnetic fields at finite baryon density.
\end{abstract}

\maketitle

%------------------------------------------------------------------------------------
%       introduction
%------------------------------------------------------------------------------------
\section{Introduction}

The macroscopic properties of a system in thermodynamic equilibrium are fundamentally characterized by its equation of state (EoS), which encapsulates the relationships among key bulk observables such as pressure, energy density, and entropy density. In the context of strongly interacting matter governed by quantum chromodynamics (QCD), the EoS serves as a cornerstone for understanding the transition between distinct regimes---from hadronic confinement at low temperatures to asymptotic freedom at high energies---and the underlying changes in degrees of freedom. A central objective of both theoretical and phenomenological efforts is to characterize the thermodynamic response of QCD matter under varying control parameters---such as temperature, baryon chemical potential, and external magnetic fields---thereby mapping out the QCD phase structure relevant to a wide range of physical environments.

Interestingly, magnetic fields are ubiquitously present across diverse physical settings. When sufficiently strong---reaching magnitudes comparable to the characteristic interaction scales---they can profoundly alter thermodynamic properties and emergent phenomena in strongly interacting matter. For instance, in cosmology, magnetic fields generated during the electroweak phase transition can modify the EoS of the early universe plasma, thereby affecting the Friedmann equations governing cosmic expansion~\cite{Vachaspati:1991nm, Enqvist:1993np, Ichiki:2006cd, Subramanian:2015lua}. In astrophysics, especially strongly magnetized neutron stars (magnetars), both surface and core magnetic fields can impact the dense matter EoS and influence mass-radius relationships governed by the Tolman-Oppenheimer-Volkoff (TOV) equations~\cite{Duncan:1992hi, Harding:2006qn}. One of the most intriguing scenarios occurs in off-central heavy-ion collisions, where high-velocity spectator charges produce some of the strongest magnetic fields predicted, estimated to reach magnitudes on the order of $\Lambda_{\rm QCD}^2$~\cite{Skokov:2009qp,Deng:2012pc}. Such strong magnetic fields are expected to induce significant non-perturbative effects on the QCD EoS, thereby altering the hydrodynamic evolution of the produced matter. These effects can manifest as striking macroscopic phenomena in QCD matter, most prominently the chiral magnetic effect \cite{Kharzeev:2007jp,Fukushima:2008xe,Kharzeev:2020jxw}. The quest to uncover such phenomena has sparked intensive theoretical and experimental investigations~\cite{Fukushima:2009ft,Fu:2013ica,Fukushima:2016vix,STAR:2021mii,Kharzeev:2022hqz, Ding:2023bft, ALICE:2025mkk} (see~\cite{Endrodi:2024cqn,Adhikari:2024bfa} for recent reviews), establishing magnetic field effects as a vital component in the exploration of QCD under extreme conditions.

The properties of QCD matter in strong magnetic fields—including its EoS—have been explored in effective frameworks,
including the linear sigma model \cite{Mizher:2010zb,Ferrari:2012yw,Ayala:2014gwa,Tawfik:2014hwa,Ayala:2020muk,Tawfik:2021eeb,Ayala:2023llp}, Nambu–Jona-Lasinio (NJL) model (along with their various extensions) \cite{Fukushima:2010fe,Skokov:2011ib,Fu:2013ica,Miransky:2015ava,Ayala:2016bbi,Farias:2016gmy,Cao:2021rwx,Sheng:2022ssp,Chahal:2023khc,Mao:2024gox,Ali:2024mnn,Kawaguchi:2024edu,Mei:2024rjg,Mao:2025toi},  hadron resonance gas (HRG) model \cite{Endrodi:2013cs,Bhattacharyya:2015pra,Fukushima:2016vix,Ferreira:2018pux,Kadam:2019rzo,Marczenko:2024kko,Vovchenko:2024wbg,Samanta:2025mrq}, perturbation theory~\cite{Blaizot:2012sd,Karmakar:2020mnj,Fraga:2023lzn,Fraga:2025juh}, holographic QCD \cite{Rebhan:2009vc,Preis:2010cq,Fukushima:2013zga,Dudal:2018rki,Fukushima:2021got,Zhu:2023aaq}, AdS/CFT~\cite{Critelli:2014kra,Dudal:2014jfa,Yin:2021zhs,Zhu:2025edv} as well as the functional renormalizaiton group approach~\cite{Wen:2025cpq,Wen:2023qcz,Mueller:2015fka}.
These semi-analytical approaches have provided valuable insights into magnetic catalysis, pressure anisotropies, and conserved charge fluctuations, despite their reliance on model-specific assumptions. However, a quantitative and systematic understanding of the QCD EoS under strong magnetic fields necessitates non-perturbative methods, where lattice QCD plays a central role by enabling direct ab initio calculations of the underlying theory.

Early lattice QCD investigations employed chromomagnetic fields confined to the gluonic sector~\cite{Cea:2002wx}. A pivotal advancement followed with the introduction of external QED magnetic fields via modified U(1) gauge links that couple directly to electrically charged quarks~\cite{DElia:2010abb}, thereby establishing a more direct connection with physical systems such as heavy-ion collisions and magnetars. Notably, magnetic fields do not induce the sign problem that plagues lattice QCD at finite density or with electric fields. Over the past decade, lattice QCD has substantially enhanced our understanding of strong magnetic fields effects, revealing significant modifications in key QCD properties~\cite{Endrodi:2024cqn}, including QCD equation of state at vanishing baryon chemical potential~\cite{Bali:2014kia}, the phase diagram~\cite{Bali:2011qj,Braguta:2019yci,DElia:2021yvk,Ding:2020inp,DElia:2025ybj}, transport characteristics~\cite{Astrakhantsev:2019zkr,Velasco:2022gaw, Almirante:2024lqn,Brandt:2024fpc}, and in-medium hadron properties~\cite{Bonati:2015dka,Endrodi:2019whh,Ding:2022tqn,Ding:2025pbu}. Most notable among these findings are now well-known phenomena of lowering of the transition temperature and the emergence of inverse magnetic catalysis \cite{Bali:2011qj,Bali:2012zg,Bruckmann:2013oba,Bali:2013esa,Bonati:2016kxj,DElia:2018xwo,Endrodi:2019zrl,Brandt:2023dir,DElia:2011koc,Bali:2012zg,Ding:2020hxw,Ding:2025pbu,Ding:2022tqn}. Despite these advances, the impact of magnetic fields on the QCD EoS at finite baryon density remains largely unexplored, mainly because a real baryon chemical potential introduces a severe sign problem that hampers direct non-perturbative simulations.

To circumvent the sign problem at nonzero baryon density, lattice QCD typically employs controlled approaches such as Taylor expansion around $\muB=0$ and simulations at imaginary chemical potential with analytic continuation~\cite{Allton:2002zi,Gavai:2003mf,HotQCD:2012fhj,Bazavov:2020bjn,Bollweg:2021vqf,Bazavov:2017dus,Bollweg:2022fqq,Roberge:1986mm,deForcrand:2002hgr,DElia:2002tig,Clarke:2024ugt,Adam:2025phc,DElia:2016jqh}. In the framework of the Taylor expansion, the leading-order coefficients are determined by fluctuations and correlations of conserved charges—net baryon number (\tB), electric charge (\tQ), and strangeness (\tS). Along this line, Ref.~\cite{Ding:2021cwv} examined leading-order conserved-charge fluctuations at $eB>0$ for a heavier-than-physical pion mass ($M_\pi\simeq220~\mathrm{MeV}$) and a single lattice spacing.   This program was subsequently extended to the physical pion mass ($M_\pi\simeq135~{\rm MeV}$), proposing the baryon electric charge correlation as a QCD magnetometer, given its pronounced magnetic enhancements ~\cite{Ding:2022uwj, Ding:2023bft, Ding:2025siy, Ding:2025jfz}, and offering a useful probe for final-stage magnetic field effects in heavy-ion collisions~\cite{Pandav:2022xxx,Nonaka:2023xkg, ALICE:2025mkk}. These works were followed by QCD EoS study at the physical point in $(2+1+1)$-flavor, although performed at finite lattice spacing $N_\tau=8$ \cite{Borsanyi:2023buy, Borsanyi:2025mrf, MarquesValois:2025nzo} as well as recent investigations at imaginary chemical potential \cite{Astrakhantsev:2024mat}. However, a systematic determination of leading-order bulk thermodynamic coefficients in magnetized matter---capturing the thermal--magnetic interplay near the transition and spanning strangeness-neutral and charge-constrained systems---remains absent.

In this work, we present continuum-estimated $(2+1)$-flavor lattice QCD results for the leading-order Taylor expansion coefficients of the EoS in strong magnetic fields at finite baryon density. We focus on strangeness-neutral systems relevant to heavy-ion collisions, incorporating both strangeness neutrality and slight isospin asymmetry. Using recently computed leading-order conserved-charge susceptibilities~\cite{Ding:2025jfz}, we determine the $T$--$eB$ dependence of the pressure coefficient $P_2$, the baryon number density coefficient $N_1^{\tB}$, and the energy-like coefficients: $\Theta_2$ (trace anomaly coefficient), $\epsilon_2$ (energy density coefficient), and $\sigma_2$ (entropy density coefficient), for $145\lesssim T\lesssim165$ MeV and $eB\leq0.8~{\rm GeV}^2$ ($\sim45 M_\pi^2$). The results reveal non-monotonic structures absent at vanishing or weak fields. We also examine isospin-parameter dependence, contrast strangeness-neutral systems with the $\muQ=\muS=0$ case, and interpret our findings via the HRG model at low $T$ and weak $eB$, and the magnetized ideal gas at high $T$ and strong fields.

The paper is organized as follows. Sec.~\ref{sec:CCTE} outlines the Taylor expansion formalism for conserved-charge susceptibilities, the definition of leading-order coefficients, and the strangeness-neutrality condition. Sec.~\ref{sec:hrg_igl} summarizes the HRG and ideal-gas benchmarks in magnetic fields. Sec.~\ref{sec:setup} describes the lattice setup and simulation parameters. Sec.~\ref{sec:initial_cond} presents the $T$--$eB$ dependence of $\muQ/\muB$ and $\muS/\muB$ under different isospin constraints. Sec.~\ref{sec:eos} reports leading-order results for $P_2$, $N_1^{\tB}$, $\Theta_2$, $\epsilon_2$, and $\sigma_2$, for both strangeness-neutral and $\muQ=\muS=0$ systems. Finally, we summarize in Sec.~\ref{sec:summary}. In the Appendices, Appendix~\ref{app:data} compiles the lattice data results and details the $T$--$eB$ interpolation and continuum-extrapolation procedure; Appendix~\ref{app:param_eos} provides compact analytic parameterizations for $q_1(T,eB)$, $s_1(T,eB)$, and $P_2(T,eB)$; in addition, Appendix~\ref{app:param_flu} provides parameterizations for the six second-order conserved-charge fluctuations, as reported in Ref.~\cite{Ding:2025jfz}.

%%%%%%
%% TE and THMDY in CONSERVED CHARGE BASIS
%%%%%%
\section{Taylor expansion and thermodynamics in the conserved-charge basis }
\label{sec:CCTE}
The thermodynamic pressure in QCD, characterizing the bulk properties of a strongly interacting system in thermal equilibrium and their response to external parameters such as temperature ($T$), quark chemical potentials ($\hat{\mu}_f \equiv \mu_f / T$) for each flavor $f$, and magnetic field strength ($eB$) at volume $V$, can be expressed in terms of the grand canonical partition function $\mathcal{Z}$ as follows:
\beq
\hat{p} \equiv \frac{p}{T^4} = \frac{1}{VT^3} \ln \mathcal{Z}(T, V, \hat{\mu}_f, eB).
\eeq

For small quark chemical potentials, the partition function can be Taylor-expanded in powers of $\hat{\mu}_f$. Considering three flavors $f = \{u, d, s\}$, the normalized pressure takes the form \cite{Allton:2002zi,Gavai:2003mf}
\bea
\hat{p}(T,eB,\hmu_{u,d,s}) = \sum_{i,j,k=0}^{\infty} \frac{1}{i!j!k!}~\chi^{uds}_{ijk}~\hat{\mu}_u^i \hat{\mu}_d^j \hat{\mu}_s^k,
\eea
where the coefficients of the expansion, $\chi^{uds}_{ijk}\equiv\chi^{uds}_{ijk}(T,eB,V)$, denote the generalized quark number susceptibilities independent of chemical potential and defined as derivatives of the logarithm of the partition function to the chemical potentials at vanishing chemical potentials
\bea
\chi^{uds}_{ijk} =\frac{1}{VT^3} \left( \frac{\partial}{\partial  \hat{\mu}_{u}} \right)^i \left( \frac{\partial}{\partial  \hat{\mu}_{d}} \right)^j \left( \frac{\partial}{\partial  \hat{\mu}_{s}} \right)^k \ln \mathcal{Z}\Bigg|_{\hat{\mu}_{u,d,s}=0}\,.
\label{eq:suscp_uds}
\eea 
Due to the invariance of QCD under charge conjugation, reflecting the particle-antiparticle symmetry, only susceptibilities with an even sum of indices of $\chi^{uds}_{ijk}$ ($\{i,j,k\} \in \mathbb{Z}$ and $i + j + k \in 2\mathbb{Z}$) are nonzero. 

In principle, the above fundamental quark flavor basis is related to the physical conserved charges basis for strong interactions, such that \bea
\hmu_u &=& \frac{1}{3} \hmu_\tB + \frac{2}{3} \hmu_\tQ\,, \nn \\ 
\hmu_d &=& \frac{1}{3} \hmu_\tB - \frac{1}{3} \hmu_\tQ\,, \nn \\
\hmu_s &=& \frac{1}{3} \hmu_\tB - \frac{1}{3} \hmu_\tQ - \hmu_\tS\,,
\label{eq:mu_uds_BQS}
\eea
where $\hmu_{\rm B,Q,S}\equiv \mu_{\tB,\tQ,\tS}/T$ are physical basis chemical potentials associated with conserved charges: net baryon number $\tB$, electric charge $\tQ$, and strangeness $\tS$. Furthermore, with the help of the above relations, the quark flavor susceptibilities can be expressed in terms of their conserved charge counterparts and vice versa, that is 
$\chi^{uds}_{ijk} \leftrightarrow \chi^{\text{BQS}}_{ijk}$, leading to the pressure series written in physical basis as follows:
\bea
\hat{p} (T,eB,\hmu_{\rm B,Q,S})
&=& \sum_{ijk}\frac{1}{i!j!k!}~ \chi^{\tB \tQ \tS}_{ijk}~ \hat{\mu}^{i}_{\tB} \hat{\mu}^{j}_{\tQ} \hat{\mu}^{k}_{\tS}\,.
\label{eq:pBQS}
\eea
Analogous to quark number susceptibilities, the generalized susceptibilities of conserved charges can be expressed as
\bea
\label{eq:cc_suscp}
\quad \chi^{\tB \tQ \tS}_{ijk}(T,eB) = \frac{\partial^{i+j+k}}{\partial \hat{\mu}_{\tB}^{i} \partial \hat{\mu}_{\tQ}^j \partial  \hat{\mu}_{\tS}^k} \hat{p} \Bigg|_{\hat{\mu}_{\tB,\tQ,\tS}=0}\,,
\eea
where the leading-order (second-order) with $i+j+k=2$ corresponds to the fluctuations of and correlations among conserved charges\footnote{For $i+j+k=0$, the system corresponds to thermodynamics at vanishing chemical potential, that is, in the absence of net conserved charge densities}. In general, the above susceptibilities are in terms of all three conserved charges; however, for brevity, we drop the superscript when the corresponding subscript is zero. 

The conserved charge number densities, $\hat{n}^{X}\equiv \hat{n}^{X}(T,eB,\hmu_{\rm B,Q,S})$ for ${X}\in \{\rm B, Q,S \}$, are obtained as corresponding chemical potential derivative of the pressure in hot magnetized medium, and with the help of the expansion coefficients $\chi^{\tB \tQ \tS}_{ijk}$ entering the Taylor series of pressure, we have 
\bea
\hat{n}^{X} \equiv {n^X}/{T^3} &=& {\partial \hat{p}}/{\partial \hat{\mu}_{X}}\nn \\
&=& \sum_{ijk}\frac{1}{i!j!k!}~\chi^{\tB \tQ \tS}_{ijk}~ \frac{\partial}{\partial \hat{\mu}_{X}} \left( \hat{\mu}^{i}_{\tB} \hat{\mu}^{j}_{\tQ} \hat{\mu}^{k}_{\tS} \right).
\label{eq:nBQS}
\eea

To define higher-order bulk observables like energy ${\hat{\epsilon}} (T,eB, \hmu_{\tB,\tQ,\tS}) $ and entropy ${\hat{s}} (T,eB, \hmu_{\tB,\tQ,\tS}) $ densities, we need to compute temperature derivatives of the expansion coefficients entering the Taylor series of $\hat{p}$, that is, 
\beq
\Xi^{\rm BQS}_{ijk}(T,eB) = T~\frac{\partial }{\partial T}\chi^{\tB \tQ \tS}_{ijk}(T,eB)\,.
\eeq 
These derivative coefficients in turn enter the expansion of the trace anomaly through the first law of thermodynamics, and we have
\bea
\hat{\Theta}(T,eB,\hmu_{\tB,\tQ,\tS}) &\equiv& {\hat{\epsilon} - 3\hat{p}} = T \frac{\partial \hat{p}}{ \partial T} \nn \\
&=& \sum_{ijk}\frac{1}{i!j!k!}~\Xi^{\tB \tQ \tS}_{ijk}~ \hat{\mu}^{i}_{\tB} \hat{\mu}^{j}_{\tQ} \hat{\mu}^{k}_{\tS}\,,
\label{eq:trace_anomaly}
\eea
which encapsulates the deviations from conformality in QCD (non-vanishing trace of energy-momentum tensor) arising due to quantum interactions and renormalization, through temperature derivatives of pressure series coefficients. With the help of the Taylor coefficients of trace anomaly, we can write the expansion of energy density,
\bea
\label{eq:eps}
{\hat{\epsilon}} (T,eB, \hmu_{\tB,\tQ,\tS}) &\equiv& {\hat{\Theta}+3\hat{p}} \nn \\
& =& \sum_{ijk}\frac{1}{i!j!k!}~\left(\Xi^{\tB \tQ \tS}_{ijk} + 3\chi^{\tB \tQ \tS}_{ijk} \right)~\nn \\ 
&&\qquad\qquad \times~ \hat{\mu}^{i}_{\tB} \hat{\mu}^{j}_{\tQ} \hat{\mu}^{k}_{\tS} \,, 
\eea
and entropy density, following directly from the Gibbs relation,
 \bea
\label{eq:sig}
{\hat{s}} (T,eB, \hmu_{\tB,\tQ,\tS}) &\equiv& {\hat{\epsilon}+\hat{p} - \sum_{X\in\{\tB,\tQ,\tS\}} \hat{\mu}_X \hat{n}^{X}} \nn \\
& =& \sum_{ijk}\big(\Xi^{\tB \tQ \tS}_{ijk} 
+ \left[ 4-(i+j+k)\right] \chi^{\tB \tQ \tS}_{ijk} \big)
\nn \\ 
&&\qquad\qquad \times~ \frac{1}{i!j!k!}~\hat{\mu}^{i}_{\tB} \hat{\mu}^{j}_{\tQ} \hat{\mu}^{k}_{\tS}\,.
\eea
The above-defined observables highlight the non-scale-invariant nature of QCD, offering insights into strong interactions and phase transitions under varying control parameters relevant to heavy-ion collisions. We now focus on the leading-order ($i+j+k=2$) contributions to bulk thermodynamic observables, which encode the dominant thermo-magnetic response of hot magnetized QCD matter at small chemical potentials.

\subsection{Leading-order thermodynamics }
\label{sec:bulk_lo_muBQS}
Starting from \autoref{eq:pBQS}, we define the leading-order (second-order) bulk thermodynamic quantities. The leading-order thermal pressure, denoted as $\hat{p}_{\rm LO} \equiv \hat{p}_{\rm LO}(T,eB, {\hmu_{\tB,\tQ,\tS}})$, can be expressed as follows:
\bea
\label{eq:plo}
\hat{p}_{\rm LO} &=& \frac{1}{2!}\left( \chi^{\tB}_{2} \hmu_{\tB}^2 + \chi^{\tQ}_{2} \hmu_{\tQ}^2 +  \chi^{\tS}_{2} \hmu_{\tS}^2 \right) \nn \\
&& \quad +~ \chi^{\tB \tQ}_{11} \hmu_{\tB} \hmu_{\tQ} + \chi^{\tB \tS}_{11} \hmu_{\tB} \hmu_{\tS} + \chi^{\tQ \tS}_{11} \hmu_{\tQ} \hmu_{\tS} \nn \\
&=&  \hf \hat{\boldsymbol \mu}^{\rm T}~ {\boldsymbol \chi}^{\rm BQS}_{\rm LO}~ \hat{\boldsymbol \mu}\,,
\eea
where ${\boldsymbol \chi}^{\rm BQS}_{\rm LO} \equiv {\boldsymbol \chi}^{\rm BQS}_{\rm LO}(T,eB)$ denotes the susceptibilities for the leading order,
\bea
{\boldsymbol \chi}^{\rm BQS}_{\rm LO}&\equiv&  \begin{pmatrix}
\chiB & \chiBQ & \chiBS \\
\chiBQ & \chiQ & \chiQS \\
\chiBS & \chiQS & \chiS 
\end{pmatrix},\quad \hat{\boldsymbol \mu} \equiv  \begin{pmatrix}
\hat{\mu}_{\tB}  \\
\hat{\mu}_{\tQ} \\
\hat{\mu}_{\tS} 
\end{pmatrix}.
\eea
Similarly, from \autoref{eq:nBQS}, we can express the leading-order conserved charge number densities $ \hat{n}^{X}_{\rm LO} \equiv \hat{n}^{X}_{\rm LO} (T,eB,\hmu_{\tB,\tQ,\tS})$ for $X\in\{\tB,\tQ,\tS\}$ as
\bea
 \hat{n}^{\tB}_{\rm LO} 
 &=&\chi^{\tB}_{2} \hmuB  + \chi^{\tB \tQ}_{11} \hmuQ + \chi^{\tB \tS}_{11}  \hmuS, \\
 \hat{n}^{\tQ}_{\rm LO}
 &=& \chi^{\tB \tQ}_{11} \hmuB + \chi^{\tQ}_{2} \hmuQ   + \chi^{\tQ \tS}_{11}  \hmuS, \\
 \hat{n}^{\tS}_{\rm LO} 
 &=&  \chi^{\tB \tS}_{11} \hmuB + \chi^{\tQ \tS}_{11}   \hmuQ +\chi^{\tS}_{2} \hmuS, 
\eea
respectively. This in shorthand matrix representation reduces to
\beq
{ \hat{\boldsymbol n}_{\rm LO}} = { {\boldsymbol \chi}^{\rm BQS}_{\rm LO}~ \hat{\boldsymbol \mu}}~; \quad { \hat{\boldsymbol n}_{\rm LO}}\equiv \begin{pmatrix}
 \hat{n}^{\tB}_{\rm LO}  \\
 \hat{n}^{\tQ}_{\rm LO} \\
 \hat{n}^{\tS}_{\rm LO}
\end{pmatrix}.
\label{eq:nBQS_lo}
\eeq

Extending to the leading order for the trace anomaly $\hat{\Theta}_{\rm LO} \equiv \hat{\Theta}_{\rm LO} (T,eB,\hmu_{\tB,\tQ,\tS})$, energy density $\hat{\epsilon}_{\rm LO} \equiv \hat{\epsilon}_{\rm LO} (T,eB,\hmu_{\tB,\tQ,\tS})$, and entropy density $\hat{s}_{\rm LO} \equiv \hat{s}_{\rm LO} (T,eB,\hmu_{\tB,\tQ,\tS})$ from Eqs.~\ref{eq:trace_anomaly},~\ref{eq:eps} and~\ref{eq:sig}, we have the following relations:
\bea
\hat{\Theta}_{\rm LO} =  \hf \hat{\boldsymbol \mu}^{\rm T}~{\boldsymbol \Xi}^{\rm BQS}_{\rm LO} ~\hat{\boldsymbol \mu}, &\\
\hat{\epsilon}_{\rm LO} =  \hf \hat{\boldsymbol \mu}^{\rm T}\left( {\boldsymbol \Xi}^{\rm BQS}_{\rm LO}+3{\boldsymbol \chi}^{\rm BQS}_{\rm LO} \right) \hat{\boldsymbol \mu}, &\\
\hat{ s}_{\rm LO} =  \hf \hat{\boldsymbol\mu}^{\rm T}\left( {\boldsymbol \Xi}^{\rm BQS}_{\rm LO}+2{\boldsymbol \chi}^{\rm BQS}_{\rm LO} \right)\hat{\boldsymbol \mu}, &
\eea
respectively, where ${\boldsymbol \Xi}^{\rm BQS}_{\rm LO} \equiv {\boldsymbol \Xi}^{\rm BQS}_{\rm LO} (T,eB)$ denotes the temperature derivative of susceptibilities of the leading order,
\beq
{\boldsymbol \Xi}^{\rm BQS}_{\rm LO}\equiv T \frac{\partial}{\partial T}  {\boldsymbol \chi}^{\rm BQS}_{\rm LO}.
\eeq
Even the leading-order Taylor expansion of the EoS above has several free control parameters, mainly the physical conserved charge potentials. However, the chemical potentials are interrelated and not independent; they are constrained by physical conditions specific to the system under study. To simplify the analysis while preserving essential characteristics of QCD matter, one often considers scenarios with vanishing electric charge and strangeness chemical potentials. Particularly important for modeling strongly interacting matter created in heavy-ion collisions is the condition of strangeness neutrality. We outline the theoretical framework corresponding to this physically relevant scenario in the next subsection.

\subsection{Strangeness-neutral systems}
\label{sec:ns0-sys}
In heavy-ion collision experiments, the colliding nuclei are initially net-strangeness neutral, and their valence quark content constrains the conserved charges in the baryon number, electric charge, and strangeness sector. Following this physical premise, lattice QCD studies consistently incorporate corresponding constraints, such as strangeness neutrality and isospin asymmetry (characterized by charge-to-baryon ratio) of the colliding nuclei, on the conserved charge chemical potentials \cite{Bazavov:2012vg, Bazavov:2014xya, Bollweg:2021vqf}. Assuming that the fluctuations are measured from a thermalized subvolume in local equilibrium within a specific experimental acceptance window, the following constraints are applied to reflect the initial strangeness neutrality and isospin asymmetry of the colliding nuclei:
\bea
 \label{eq:strange-neutral_isospin}
 \hat{n}^{\tS} = 0,\quad  {n}^{\tQ} /  {n}^{\tB} = r.
\eea
Here $r$ denotes the isospin parameter, which characterizes the charge-to-baryon ratio of various colliding nuclei systems. The corresponding values of $r\in [0,1]$ span different regimes of isospin asymmetry:
\begin{itemize}
    \item[--] \textit{Symmetric system} ($r=0.5$): This scenario corresponds to colliding systems with effectively equal numbers of protons and neutrons, and might be of relevance to oxygen-oxygen ($O$ atom-type) collisions. Notably, in the absence of magnetic fields ($eB=0$), the electric charge chemical potential $\hmu_{\rm Q}$ vanishes at all orders. However, the presence of a magnetic field induces isospin-breaking effects leading to nonzero $\hmuQ$.
    
    \item[--] \textit{Slight asymmetric systems} ($r\simeq0.4$):   These systems are particularly relevant for heavy-ion collisions involving nuclei such as {\it Pb/Au/Zr/Ru} atoms, which exhibit a moderate isospin imbalance due to an excess of neutrons. For instance, in collisions involving the nuclei $^{208}_{82}Pb$ and $^{197}_{79}Au$, with valence quark content $u:d:s \equiv 46.5:53.5:0$, the isospin parameter can be estimated as
    \bea
    \label{eq:PbAu_constraint}
    r = \frac{ {n}^{\tQ}}{  {n}^{\tB}} = \frac{\frac{2}{3} \times 0.465 -\frac{1}{3} \times0.535 }{\frac{1}{3}\times 0.465 +\frac{1}{3}\times 0.535} &\simeq& 0.4 \,.
    \eea
    For other heavy atoms, $^{96}_{40}Zr$ and $^{96}_{44}Ru$, the parameter $r$ can be estimated as $r=0.417$ and $r=0.458$, respectively.

    \item[--] \textit{Maximally asymmetric neutral system} ($r=0.0$):  This extreme case corresponds to a system composed of purely electrically neutral baryonic matter, such as all neutrons with $\hat{n}^{\rm Q} = 0$ and thereby $r=0$, resulting in a maximal isospin asymmetry that is electrically neutral.

    \item[--] \textit{Maximally asymmetric charged system}  ($r=1.0$):  This extreme case corresponds to a system composed of all positively charged baryons, such as composed entirely of protons with $\hat{n}^{\rm Q} =\hat{n}^{\rm B}$ and thereby $r\equiv n^{\tQ}/n^{\tB} = 1$, resulting in maximally isospin asymmetry that is electrically charged.  
\end{itemize}

To implement these constraints in QCD thermodynamics, we first Taylor expand the electric charge and strangeness chemical potentials in powers of the baryon chemical potential:
\bea
\label{eq:muQ}
 \hat{\mu}_{\tQ} &\equiv& \hat{\mu}_{\tQ}(T,eB,\hat{\mu}_{\tB}) \nn \\
 &=& \sum_{k=1}^{\infty} q_{2k-1}\hat{\mu}_{\tB}^{2k-1} \nn \\
&=& q_1 (T,eB) \hat{\mu}_{\tB} + q_3 (T,eB) \hat{\mu}_{\tB}^3+ \mathcal{O}(\hat{\mu}_{\tB}^5),
\eea
and
\bea
\label{eq:muS}
\hat{\mu}_{\tS} &\equiv& \hat{\mu}_{\tS}(T,eB,\hat{\mu}_{\tB}) \nn \\  &=& \sum_{k=1}^{\infty} s_{2k-1}\hat{\mu}_{\tB}^{2k-1} \nn \\
&=& s_1 (T,eB) \hat{\mu}_{\tB} + s_3 (T,eB) \hat{\mu}_{\tB}^3+ \mathcal{O}(\hat{\mu}_{\tB}^5).
\eea
The expansion coefficients $q_{2k-1}$ and $s_{2k-1}$ depend on temperature and magnetic field strength, and they can be computed order by order from the strangeness neutrality and isospin asymmetry constraints. 
The leading-order coefficients $q_1(T,eB)\equiv \left({\mu_\tQ}\big/{\mu_\tB} \right)_{\rm LO}$ and $s_1(T,eB)\equiv\left({\mu_\tS}\big/{\mu_\tB} \right)_{\rm LO}$ can be computed using the leading order number densities expressed in terms of baryon chemical potential. The leading-order baryon number density from~\autoref{eq:nBQS_lo} can be expressed purely in terms of baryon chemical potential as \cite{Bazavov:2012vg}
\bea
    \hat{n}^{\tB}_{\rm LO} (T, eB,\hmuB)&=&\chi^{\tB}_{2} \hmuB  + \chi^{\tB \tQ}_{11} \hmuQ + \chi^{\tB \tS}_{11}  \hmuS\nn \\
    &=& \left(\chi^{\tB}_{2} + q_1\chi^{\tB \tQ}_{11} + s_1\chi^{\tB \tS}_{11} \right)  \hat{\mu}_{\tB},
\label{eq:nB_LO}
\eea
and similarly for electric charge and strangeness number densities,
\bea
 \hat{n}^{\tQ}_{\rm LO}(T, eB,\hmuB)&=& \left(q_1\chi^{\tQ}_{2} + \chi^{\tB \tQ}_{11} + s_1\chi^{\tQ \tS}_{11} \right)  \hat{\mu}_{\tB},  \\ 
 \hat{n}^{\tS}_{\rm LO}(T, eB,\hmuB)&=& \left(s_1\chi^{\tS}_{2} + \chi^{\tB \tS}_{11} + q_1\chi^{\tQ \tS}_{11} \right)  \hat{\mu}_{\tB},
\eea
respectively. Using these leading-order number densities, we can solve the initial nuclei constraint mentioned in~\autoref{eq:strange-neutral_isospin} at the leading order, that is,
\bea
\hat{n}^{\tS}_{\rm LO}= 0 ~\xrightarrow{}&&s_1\chi^{\tS}_{2} + \chi^{\tB \tS}_{11} + q_1\chi^{\tQ \tS}_{11}=0 \,, \\
 \hat{n}^{\tQ}_{\rm LO}/ \hat{n}^{\tB}_{\rm LO}= r~ \xrightarrow{}&& q_1\chi^{\tQ}_{2} + \chi^{\tB \tQ}_{11} + s_1\chi^{\tQ \tS}_{11} \nn \\
 &&\quad = r \left( \chi^{\tB}_{2} + q_1\chi^{\tB \tQ}_{11} + s_1\chi^{\tB \tS}_{11} \right) \,,
\eea
which leads us to the expressions for $q_1\equiv q_1(T,eB)$ and $s_1\equiv s_1(T,eB)$~\cite{Bazavov:2012vg},
\bea
q_1 = \frac{r \left( \chi_{2}^{\tB}  \chi_{2}^{\tS} - \chi_{11}^{\tB \tS} \chi_{11}^{\tB \tS} \right) - \left(\chi_{11}^{\tB \tQ} \chi_{2}^{\tS} - \chi_{11}^{\tB \tS} \chi_{11}^{\tQ \tS} \right)}{\left(\chi_{2}^{\tQ } \chi_{2}^{\tS} - \chi_{11}^{\tQ \tS}  \chi_{11}^{\tQ \tS}\right) - r\left( \chi_{11}^{\tB \tQ} \chi_{2}^{\tS} -\chi_{11}^{\tB \tS} \chi_{11}^{\tQ \tS} \right)},
\label{eq:q1_def}\\
s_1 = - {\left( \chi_{11}^{\tB \tS} + q_1 \chi_{11}^{\tQ \tS}\right)}/ { \chi_{2}^{\tS}},
\label{eq:s1_def}
\eea
as explicit functions of the fluctuations of conserved charges and the isospin parameter $r$. In our recent work, we computed the second-order fluctuations and correlations of conserved charges and systematically studied their temperature and magnetic field dependence \cite{Ding:2025jfz}. Building on these continuum-estimated fluctuation results, we will examine in Sec.~\ref{sec:initial_cond} the leading-order coefficients of chemical potential ratios, $q_1$ and $s_1$, in strangeness-neutral systems across a range of isospin conditions. 
These coefficients, in turn, govern the expansion of bulk thermodynamic observables under strangeness-neutral constraints.

\subsection{$\hmuB$-series expansion for bulk thermodynamics}
\label{sec:muB_expansion}

Earlier in this section, we established that the bulk observables, such as $\hat{p}(T,eB,\hmu_{\tB,\tQ,\tS})$ in~\autoref{eq:pBQS}, evaluated using conserved charge susceptibilities, have several free control parameters, mainly the three interrelated chemical potentials. We effectively utilize strangeness neutrality and isospin asymmetry conditions discussed in previous Sec.~\ref{sec:ns0-sys} to help constrain this pressure Taylor expansion purely as a baryon chemical potential series expansion $\hat{p}:\hat{p}(T,eB,\hmu_{\tB,\tQ,\tS})\to \hat{p}(T,eB,\hmu_{\tB})$, that is,
\bea
\label{eq:p_q1s1}
\hat{p} (T, eB,\hmuB)
&=& \sum_{ijk}\frac{1}{i!j!k!}~ \chi^{\tB \tQ \tS}_{ijk}~ \hat{\mu}^{i}_{\tB} \nn \\
&&\quad \times~ \left(  q_1 \hat{\mu}_{\tB} + \mathcal{O}(\hat{\mu}_{\tB}^3) \right)^{j}\quad \nn \\
&&\quad \times~ \left(  s_1  \hat{\mu}_{\tB} + \mathcal{O}(\hat{\mu}_{\tB}^3) \right)^{k}.
\eea
This constrained relation leads to a series expansion purely in terms of $\hmuB$, that is
\bea
\Delta\hat{p}&\equiv& \hat{p}(T,eB,\hmu_{\tB}) -\hat{p}(T,eB,0) \nn \\
&=& \sum_{k=1}^{\infty} P_{2k}(T,eB) \hat{\mu}_{\tB}^{2k}\,,
\label{eq:p2k}
\eea
where $\Delta\hat{p}$ denotes the pressure difference encoding medium induced thermo-magnetic effects at finite baryon chemical potential. $P_{2k}$ are the Taylor expansion coefficients and are explicit functions of the generalized conserved charge susceptibilities and chemical potential ratios, that is $P_{2k}(T,eB)\equiv f(\chi^{\rm BQS}_{ijk}, q_{2k-1},s_{2k-1})$. $\hat{p}(T,eB,0) $ often termed as $P_0\equiv \hat{p}(T,eB)=\chi^{\tB \tQ \tS}_{000}$, represents the thermal and quantum effects including vacuum fluctuations, of pressure at vanishing baryon density \footnote{For more details, see Ref.~\cite{Bali:2020bcn}, where the weak-field behavior of $P_0$ is characterized in terms of the QCD magnetic susceptibility.}.

Building upon the baryon chemical potential expansion of pressure mentioned above, we can define baryon number density by taking derivatives with respect to $\hmuB$, yielding a Taylor series expansion in odd powers of $\hmuB$: 
\beq
\hat{n}^{\rm B} (T,eB,\hmuB) = \sum_{k=1}^{\infty} N^{\rm B}_{2k-1}(T,eB) \hat{\mu}_{\tB}^{2k-1}.
\label{eq:nB_muB}
\eeq
Here, $N^{\rm B}_{2k-1}$ are the expansion coefficients of baryon number density, with the dominant leading-order contribution given by $N^{\rm B}_{1}$. Note that, $N^{\rm B}_{1}\hat{\mu}_{\tB}$ is nothing but \autoref{eq:nB_LO}. Since both $N^{\rm B}_{2k-1}$ and $P_{2k}$ arise from the same Taylor expansion, they are related in the following manner:
\bea
P_{2k} &=& \frac{1}{2k} \Big[N^{\rm B}_{2k-1} \nn \\
&&\qquad +~r\sum_{j=1}^{k}(2j-1)~q_{2j-1} N^{\rm B}_{2k -2j+1}  \Big].
\label{eq:ratio_NBk_P2k}
\eea

Extending the framework applied for the pressure and number density expansions, incorporating strangeness neutrality and isospin asymmetry constraint, the expansion of trace anomaly, energy density, and entropy density, introduced in the Eqs.~\ref{eq:trace_anomaly},~\ref{eq:eps} and~\ref{eq:sig}, can be systematically expanded in powers of $\hmuB$ as,
\bea
\label{eq:theta2k_muB}
\Delta \hat{\Theta}&\equiv& \hat{\Theta} (T,eB, \hmuB) -  \hat{\Theta} (T,eB, 0) \nn \\ &=& \sum_{k=1}^{\infty}  \Theta_{2k}(T,eB) ~\hmuB^{2k},\\
\label{eq:eps2k_muB}
\Delta \hat{\epsilon}&\equiv& \hat{\epsilon} (T,eB, \hmuB) -  \hat{\epsilon} (T,eB, 0) \nn \\ &=& \sum_{k=1}^{\infty}  \epsilon_{2k}(T,eB) ~\hmuB^{2k},\\
\label{eq:sig2k_muB}
\Delta \hat{s}&\equiv& \hat{s} (T,eB, \hmuB) -  \hat{s} (T,eB, 0) \nn \\ 
&=& \sum_{k=1}^{\infty} \sigma_{2k}(T,eB)~ \hmuB^{2k},
\eea
respectively, where $\Theta_{2k}$, $\epsilon_{2k}$ and $\sigma_{2k}$ are the respective Taylor expansion coefficients.
These coefficients are constructed from specific combinations of the pressure expansion coefficients and their temperature derivatives, encapsulating the response of bulk thermodynamic quantities to finite baryon density in a hot magnetized QCD medium. In Sec.~\ref{sec:eos}, we will examine the temperature and magnetic field dependence of the dominant leading-order coefficients $P_2,~N^{\tB}_1, ~\Theta_2,~\epsilon_2$ and $\sigma_2$.

%%%%%%
% Sec: HRG IGL
%%%%%%
\section{Hadron resonance gas and ideal gas limits}
\label{sec:hrg_igl}

Lattice QCD provides a robust first-principles framework for QCD; nevertheless, certain simplistic phenomenological models can be effectively utilized to gain physical insights from lattice results. In these phenomenological descriptions of QCD matter, the relevant degrees of freedom and their interactions can be identified based on different control parameter regimes. In this section, we outline the hadron resonance gas (HRG) model at low temperatures and relatively weak magnetic fields, and the magnetized ideal gas model at high temperatures, to characterize different limiting behaviors of QCD matter.

\subsection{Low-temperature HRG at relatively weak magnetic fields}
The low-temperature confined phase of QCD can be effectively described as a medium composed of hadrons and their resonances. The HRG model provides a well-established and straightforward approach to probe the thermodynamics of hadronic QCD matter in the presence of nonzero chemical potentials \cite{Karsch:2003zq,Andronic:2005yp,Huovinen:2009yb,Vovchenko:2016ebv}. Recently, it has garnered attention to study thermodynamics in the presence of external magnetic fields, which are expected to introduce complex thermomagnetic 
effects due to modifications in the hadron spectrum \cite{Endrodi:2013cs,Bhattacharyya:2015pra,Fukushima:2016vix,Kadam:2019rzo,Ding:2021cwv,Ding:2023bft,Vovchenko:2024wbg,Marczenko:2024kko,Ding:2025jfz,Samanta:2025mrq}. With the fundamental assumption that the medium consists of non-interacting point-like hadrons and resonance states~\footnote{In principle, hadrons are not point-like particles but bear non-trivial internal structure}, the HRG pressure is expressed as the sum over individual pressure contributions $p_R$ from each resonance $R$: 
\begin{equation}
    \frac{p^{\rm HRG}}{T^4}=\frac{1}{T^4} \sum_R p_R=\frac{1}{V T^3} \sum_R \ln {\mathcal{Z}}_R(V, T, {\mu}_R, eB)\,,
    \label{eq:pressure_HRG}
\end{equation}
where ${\mathcal{Z}}_R$ denotes the contribution of resonance $R$ to the total grand canonical partition function, typically represented as an integral over momentum space $\bm{p}$, 
\begin{equation}
    \ln{\mathcal{Z}}_R= \pm V g_R \int \frac{\mathrm{~d}^3 \bm{p}}{(2 \pi)^3} \ln \left[1 \pm e^{-\left(E_R-\mu_R\right) / T}\right]\,.
    \label{eq:part_fun_HRG}
\end{equation}
Here, $T, V, eB$ are the temperature, volume, and magnetic field strength, respectively. For each resonance $R$,  the '$\pm$' sign accounts for baryons or mesons, $g_R$ is the degeneracy factor, $E_R$ is the energy, and $\mu_R= \mu_{\rm B}\tB_R + \mu_{\rm Q}\tQ_R+ \mu_{\rm S}\tS_R$ is the chemical potential, expressed in terms of the baryon number $\tB_R$, electric charge $\tQ_R$, and strangeness $\tS_R$ and their associated chemical potentials. The momentum integral can be evaluated analytically for the thermal pressure (neglecting vacuum energy terms) of neutral $p_{n,R}$ and charged $p_{c,R}$ resonances. 

 The total HRG thermal pressure can be obtained as a sum over partial pressures for all neutral and charged resonances. For neutral resonances, neglecting magnetic field effects, the energy dispersion is given by the standard relativistic relation $E_R=\sqrt{{\bm p}^2+m_R^2}$, where $m_R$ is the resonance mass. In this case, thermal pressure from neutral resonances can be expressed as~\cite{Hagedorn:1965st,Karsch:2003zq,HotQCD:2012fhj}   
\bea
\label{eq:nR_eB_hrg}
\frac{p_{n,R}}{T^4} = \frac{g_{R} m_{R}^2}{2(\pi T)^2} \sum_{k=1}^{\infty} \left(\pm 1\right)^{k+1} \frac{e^{k{\mu_{{R}}}/T}}{k^2}  \text{K}_2 \left( \frac{k m_{{R}}}{T} \right)\,,
\eea
where $k$ is the series index and $\mathrm{K}_2$ is the second-order modified Bessel function of second kind. 

In contrast, the thermal pressure from charged hadrons is significantly modified in the presence of a magnetic field due to Landau quantization. Here, discrete Landau energy levels $\epsilon_l$ are given by \cite{Endrodi:2013cs,Fukushima:2016vix, Ding:2021cwv}
\beq
\label{eq:disp_rel}
\epsilon_l  = \sqrt{m_R^2 + 2 |q_R| B (l+1/2-s_z)}\,,
\eeq
where $l$ labels Landau energy levels, $s_z$ denotes the spin projection along the magnetic field, and $q_R$ is the electric charge of the resonance. Assuming a point-like resonance structure, the thermal pressure from charged hadrons in the presence of magnetic fields can be expressed as,
\bea
\label{eq:cR_eB_hrg}
\frac{p_{c,R}}{T^4} &=& \frac{|q_R|B}{2\pi^2 T^3} \sum_{s_z =-s_R}^{s_R}   \sum_{l=0}^{\infty}     \epsilon_l \sum_{k=1}^{\infty} \nn \\ 
&&\qquad ~ \times \left(\pm 1\right)^{k+1} \frac{e^{k{\mu_R}/T}}{k}  {\rm K}_1 \left( \frac{k \epsilon_l}{T} \right)\,,
\eea
where $\mathrm{K}_1$ denotes the first-order modified Bessel function of the second kind. Here, each Landau level $\epsilon_l$ carries a degeneracy factor of $|q_R|B/2\pi$, appearing explicitly in ~\autoref{eq:cR_eB_hrg} and originating from the magnetic field-modified transverse phase space density of states:
\beq
\label{eq:dos_eB}
    g_R\int \frac{d^3 \bm{p}}{(2\pi)^3} \rightarrow \frac{|q_R|B}{2\pi} \sum_{s_z=-s_R}^{s_R} \sum_{l=0}^{\infty} \int \frac{dp_z}{2\pi}.
\eeq
In~\autoref{eq:cR_eB_hrg}, the longitudinal momentum $p_z$ in the magnetic field direction has been integrated out, while the discrete transverse momentum states result from the quantization imposed by the magnetic field. Importantly, the Landau level degeneracy and the reduction of Landau level energies for higher spin projections $s_z \geq \frac{1}{2}$, reorganize the occupation of quantum states, thus substantially impacting the thermodynamic properties of charged resonances.

From the thermal pressure derived above, generalized susceptibilities and thereby the EoS can be obtained using \autoref{eq:cc_suscp}. At vanishing chemical potentials, the fluctuations of and correlations among conserved charges for the charged resonance $R$  take the form \cite{Fukushima:2016vix, Ding:2021cwv, Ding:2023bft, Ding:2025jfz}, 
\bea
\chi_{2,R}^{X} &=& \frac{\partial^{2} \left( p_{c,R}/T^4 \right) }{\partial \left( \mu_{X}/T \right)^2}\Bigg|_{{\bm \mu} = 0} \nn \\
&=& \frac{|q_R|B}{2\pi^2 T^3}~   X_R^2 \sum_{s_z =-s_R}^{s_R}   \sum_{l=0}^{\infty}   f({\epsilon_l}), \\
\chi_{11,R}^{XY} &=& \frac{\partial^{2} \left( p_{c,R}/T^4 \right) }{\partial \left( \mu_X/T \right) \partial \left( \mu_Y/T \right)} \Bigg|_{\bm \mu = 0}\nn \\
&=& \frac{|q_R|B}{2\pi^2 T^3} ~   X_R Y_R\sum_{s_z =-s_R}^{s_R}   \sum_{l=0}^{\infty} f({\epsilon_l})  \,,
\eea
where, $X_R, Y_R$ represents the conserved charges $\{\rm B, ~Q, ~S\}$ carried by resonance $R$ and the weight factor is given by $
f(\epsilon_l) = \epsilon_l \sum_{k} (\pm)^{k+1}~ k {\mathrm{K}}_1 \left( {k\epsilon_l}/{T} \right)$. 

It is important to note that the above HRG representation is valid only for relatively weak magnetic fields. For sufficiently strong fields, the energy corresponding to the lowest Landau level for high-spin resonances, such as spin-1 mesons and spin-$3/2$ baryons, can become complex, indicating a breakdown of the HRG approximation. 

The total pressure and generalized susceptibilities depend on the set of considered resonances. In the literature, models incorporating only experimentally observed resonances, as compiled by the Particle Data Group (PDG) \cite{ParticleDataGroup:2020ssz}, are referred to as PDG-HRG, while those that also include predicted resonances from the Quark Model are known as QM-HRG. The QM-HRG includes experimentally undetected states and provides an improved description of low-temperature QCD thermodynamics in the strangeness sector by accounting for these missing resonances \cite{Alba:2017mqu, Bollweg:2021vqf}. 

Recently, within the HRG framework, experimentally relevant proxies and systematic kinematic cuts have been developed in the presence of magnetic fields \cite{ Ding:2023bft,Ding:2025jfz}, extending the zero-field case \cite{Karsch:2015zna,Bellwied:2019pxh}. Following the  Taylor expansion framework defined in the previous Sec. \ref{sec:CCTE}, bulk thermodynamics observables and EoS can thus be consistently formulated within the HRG approach.

\subsection{High-temperature magnetized ideal gas model}
At high temperatures, the fundamental degrees of freedom in QCD can be approximated as a non-interacting ideal gas, commonly referred to as the quark-gluon gas or the free limit. In the presence of an external magnetic field, the QCD pressure of magnetized ideal gas with three massless quark flavors is given by \cite{Kapusta:2006pm, Laine:2016hma, Ding:2021cwv, Ding:2025jfz}, 
\beq
\label{eq:ideal}
\frac{p^{\rm ideal}}{T^4} = \frac{8\pi^2}{45} + \sum_{f=\{u,d,s\}} \frac{3 \left| q_f\right| B}{\pi^2 T^2} \left[ \frac{\pi^2}{12} + \frac{\hat{\mu}^2_f}{4} +p_f(B) \right]\,,
\eeq
where $q_f$ and $\hat{\mu}_f\equiv{\mu}_f/T $ are the electric charge and chemical potential for flavor $ f$. The first term, ${8\pi^2}/{45}$, represents the gluon contributions and remains immune to magnetic effects. In contrast, the quark contributions are modified in external magnetic fields due to Landau quantization, leading to discrete Landau level contributions. Specifically, the term proportional to ${\hat{\mu}^2_f}$ accounts for the lowest Landau level, while the function $p_f(B)$ encodes the contributions from the higher Landau levels:
\bea
p_f(B) &=& 2\frac{\sqrt{2 \left| q_f \right| B  }}{T} \sum_{l=1}^{\infty} \sqrt{l} \sum_{k=1}^{\infty} \frac{(-1)^{k+1}}{k} ~\cosh(k \hat{\mu}_f)
\nn \\
&&\qquad \qquad \times~ {\rm K}_1\left( \frac{k\sqrt{2\left| q_f \right| Bl }}{T} \right)\,,
\eea
where $l$ is the Landau level and $k$ is the summation index for the series expansion. From the above-evaluated ideal gas pressure, the generalized susceptibilities can be obtained using \autoref{eq:suscp_uds} in the fundamental flavor basis and henceforth, in the conserved charged basis via the relations in \autoref{eq:mu_uds_BQS}. As discussed in Sec. \ref{sec:CCTE}, these susceptibilities enable the construction of leading-order coefficients for bulk thermodynamic observables.

In the context of the ideal gas approximation discussed above, one can further investigate the magnetized free limit, defined as $\sqrt{eB}/T \to \infty$ for $T\to \infty$. For the asymptotic behavior of fluctuations of and correlations among conserved charges in this limit, see Ref. \cite{Ding:2021cwv, Ding:2025jfz}. It is important to note that while the gluon contributions remain unaffected by the magnetic field and thus do not play any role in susceptibilities or bulk coefficients,  the dominant quark-sector contributions exhibit a linear dependence on magnetic fields. Furthermore, for leading-order thermodynamics ratio observables such as $q_1$, $s_1$, and $N^{\tB}_1/2P_2$, the explicit    
$eB$-dependence cancels out, resulting in saturation to finite values. These asymptotic values depend on system-specific constraints, such as the isospin parameter $r$ in strangeness-neutral systems:
\bea
q_{1,r}(\sqrt{eB}/T \to \infty) &=& -\left(\frac{1-r}{3-r}\right) \,,\\
s_{1,r}(\sqrt{eB}/T \to \infty) &=& \frac{1 -q_{1,r}({\sqrt{eB}}/T \to \infty) }{3} \nn\\
&=& \frac{2}{3}\left(\frac{2-r}{3-r} \right)\,,\\
\frac{N^{\tB}_{1,r}}{2P_{2,r}}(\sqrt{eB}/T \to \infty) &=& \frac{1}{1 + rq_{1,r}(\sqrt{eB}/T \to \infty) } \nn \\
&=& \frac{3-r}{3+r^2-2r}\,.
\eea
For reference to magnetic field-induced effects, we also note the high-temperature limit at vanishing magnetic fields, that is, $eB=0, T\to \infty$ \cite{HotQCD:2012fhj, Bazavov:2017dus}, where the leading-order thermodynamics ratios attain the following saturation:
\bea
q_{1,r}({eB=0},T \to \infty) &=& -\left(\frac{1-2r}{5-r}\right)\,,\\
s_{1,r}({eB=0},T \to \infty) &=& \frac{1 -q_{1,r}({eB=0},T \to \infty) }{3} \nn\\
&=& \frac{2-r}{5-r}\,,\\
\frac{N^{\tB}_{1,r}}{2P_{2,r}}({eB=0},T \to \infty) &=& \frac{1}{1 + rq_{1,r}({eB=0},T \to \infty)} \nn \\
&=&\frac{5-r}{5+2r^2-2r}\,.
\eea
\begin{table}[!htbp]
    \centering
    \begin{tabular}{c|c|c|c}
        \hline \hline
        ~ & $\quad r\quad$ &  $eB=0$      & $\sqrt{eB}/T \to \infty$         \\
        \hline 
        $q_{1,r}$ 
         & 0.4 & -0.0435 & -0.2308 \\
         & 0.0 & -1/5    & -1/3 \\
         & 0.5 & 0       & -1/5 \\
         & 1.0 & 1/4     & 0 \\
        \hline
        $s_{1,r}$ 
         & 0.4 & 0.3478 & 0.4103 \\
         & 0.0 & 2/5    & 4/9 \\
         & 0.5 & 1/3    & 2/5 \\
         & 1.0 & 1/4    & 1/3 \\
        \hline
        $\frac{N^{\tB}_{1,r}}{2P_{2,r}}$ 
         & 0.4 &  1.0177  &  1.1017 \\
         & 0.0 &  1       & 1 \\
         & 0.5 &  1       & 1/9 \\
         & 1.0 &  4/5     & 1 \\
        \hline \hline	
    \end{tabular}
    \caption{High-temperature free-limit $(T\to \infty)$ values of leading-order thermodynamics ratio observables, $q_{1}$, $s_{1}$ and ${N^{\tB}_1}/2P_2$ in a hot magnetized medium with $\sqrt{eB}/T\rightarrow \infty$\cite{Ding:2021cwv, Ding:2025jfz} and in the absence of magnetic fields ${eB=0}$ \cite{HotQCD:2012fhj,Bazavov:2017dus}, for strangeness-neutral systems spanning different isospin symmetry regimes characterized by the parameter $r$.}
    \label{tab:free_limit}
\end{table}

These asymptotic free-limit values in both the magnetized and unmagnetized cases are summarized in~\autoref{tab:free_limit}, highlighting their dependence on the isospin parameter $r$ in strangeness-neutral systems. We will utilize these high-temperature limits as reference benchmarks when discussing our lattice QCD results at finite temperature and magnetic fields in Secs. \ref{sec:initial_cond} and \ref{sec:eos}.

%%%%%%
%% Sec: LATTICE SETUP
%%%%%%
\section{Lattice setup}
\label{sec:setup}

The QCD partition function is given by the functional integral
\bea
\mathcal{Z} &=&  \int \mathcal{D} U~ \prod_{f=\{u,d,s\}}\left[{\rm det} {M_f}(U,B,q_f,m_f,\mu_f) \right]^{1/4} \nn \\
&&\qquad \qquad \qquad \qquad \times ~ e^{-S_{g}(\beta, U)}\,,
\eea
where  $M_f(U,B,q_f,m_f,\mu_f)$ denotes the fermion matrix for flavor $f$ for background magnetic field $B$, quark charges $q_f$, masses $m_f$ (degenerate for light quarks, $m_u=m_d$ ) and chemical potential $\mu_f$.  We have gauge links $U \in {\rm SU(3)}$, bare lattice gauge coupling $\beta = 6/g^2$ with gauge action $S_g(\beta, U)$. 
Our simulations consider (2+1)-flavor QCD with highly improved staggered quarks (HISQ) \cite{Follana:2006rc} and a tree-level improved Symanzik gauge action. These improvements in bosonic and fermion action have been extensively used by the HotQCD collaboration \cite{Bazavov:2019www, Bollweg:2021vqf, Bollweg:2022rps, Bollweg:2024epj}. 

We consider a uniform magnetic field along the $z$ direction $\Vec{B}=(0,0, B)$ and fix the gauge corresponding to the vector potential curl, using the Landau gauge \cite{Bali:2011qj,Al-Hashimi:2008quu}, leading to fixed factors $u_\mu(n) $ of the U(1) field \cite{DElia:2010abb,Ding:2021cwv, Ding:2020hxw} expressed as
\bea
u_t(n_x,n_y,n_z,n_\tau) &=& u_z(n_x,n_y,n_z,n_\tau) \nn \\
&=& u_x(n_x\neq N_x-1,n_y,n_z,n_\tau) \nn \\
&=&1,  \nn \\
u_x(N_x-1,n_y,n_z,n_\tau) &=& 
e^{-iq_fa^2 B N_x n_y},  \nn\\
u_y(n_x,n_y,n_z,n_\tau) &=& e^{iq_fa^2 B n_x}, 
\label{eq:u1_links}
\eea
which are in turn multiplied to the SU(3) gauge link $U_\mu$. Here $N_\mu = \left(N_x, N_y, N_z, N_{\tau} \right)$ denotes the lattice size, and the coordinates are labeled by integers $n=\left(n_x,n_y,n_z \right)$, for $n_\mu = 0, \dots, N_\mu-1$ with $\mu = x,y,z,\tau$. 

In this setup, the strength of the magnetic field, $eB$, is quantized by boundary conditions and finite lattice geometry following from the Stokes theorem, leading to the quantization condition~\cite{DElia:2010abb,Ding:2023bft,Al-Hashimi:2008quu,tHooft:1979rtg}: 
\beq
eB = \frac{6\pi N_b}{N_x N_y}a^{-2}, \quad N_b \in \mathbb{Z},
\eeq
where $e$ is the elementary electric charge and $N_b$ counts the magnetic flux quanta through the $x-y$ plane with lattice spacing $a$. The electric charges of the quarks are related in the following manner: $q_d=q_s=-q_u/2=-e/3$. To satisfy flux quantization across all flavors, we consider the greatest common divisor for electric charges of quarks in our implementation, i.e., $\left|q_{d}\right|=\left|q_{s}\right|=e /3$. Furthermore, the periodic boundary conditions for U(1) links imposed for all except the x-direction (up to a gauge transformation) imply periodicity in magnetic fields with a period of $N_xN_y$, leading to an additional constraint on the flux, $0\leq N_b < {N_x N_y}/{4}$. Further details about incorporating magnetic fields with the HISQ action can be found in Refs.~\cite {Ding:2020hxw, Ding:2021cwv}.

We consider a physical strange quark mass, $m_s$, and degenerate light quark masses, $m_u=m_d=m_s/27$, which correspond to a pseudo-Goldstone physical pion mass $M_{\pi} \simeq 135 ~{\rm MeV}$ at vanishing magnetic fields. The scale-setting approach is adopted from Refs.~\cite{Bollweg:2021vqf,Bazavov:2019www}. Our simulation setup employs spatially symmetric lattices, $N_{\sigma} \equiv N_x=N_y=N_z$ with fixed spatial-to-temporal aspect ratio, $N_{\sigma}/N_{\tau}=4$.  Gauge configurations were generated using a modified version of the \texttt{SIMULATeQCD} software suite~\cite{HotQCD:2023ghu} primarily on $32^3 \times 8$ and $48^3 \times 12$ lattices for nonzero magnetic flux $N_b$. 

In the control parameter space, the magnetic field strength $eB$ reaches up to $ 0.8~{\rm GeV}^2\sim 45M_\pi^2$, corresponding to the flux $N_b\in \{1,2,3,4,6,12,24,32 \}$. For the case of $N_b=0$, we have adopted lattice QCD results from Ref.~\cite{Bollweg:2021vqf}. The temperature range is focused around the pseudo-critical temperature for the QCD transition within the considered $eB$ range, $T\in [145 - 165] ~{\rm MeV} $. To mitigate discretization effects associated with magnetic fields, it is necessary that $a^2 ~eB \ll 1$, which translates to $N_b / N_\sigma^2 \ll 1$. In practice, $N_b / N_\sigma^2 < 5\%$ is commonly adopted in the literature~\cite{DElia:2021tfb,Endrodi:2019zrl}. In our simulations, the largest magnetic flux $N_b = 32$ on a $32^3 \times 8$ lattice corresponds to a maximum $N_b / N_\sigma^2$ of about $3\%$, which remains within the accepted range, ensuring that discretization effects from the magnetic field are mild. Notably, this ratio becomes even smaller on the $48^3 \times 12$ lattice.

%%%%
\begin{figure*}[t]
\centering

\includegraphics[width=0.45\textwidth]{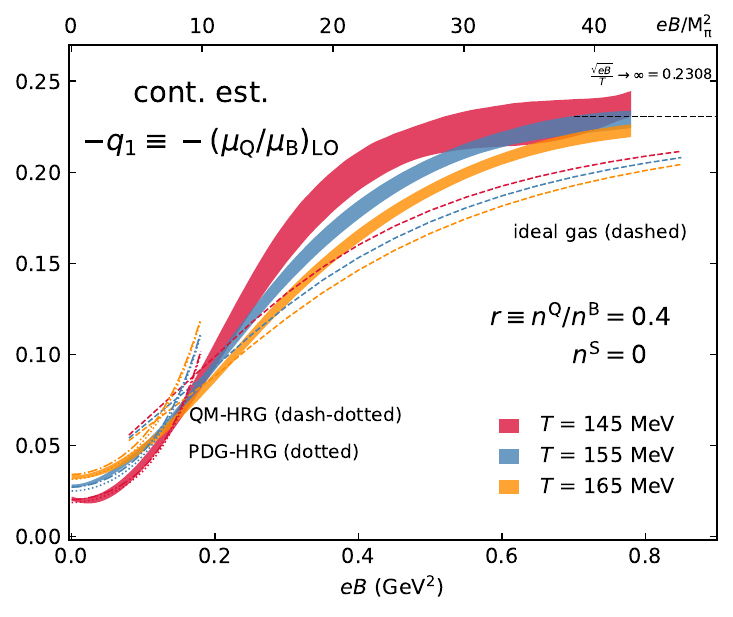}
\includegraphics[width=0.45\textwidth]{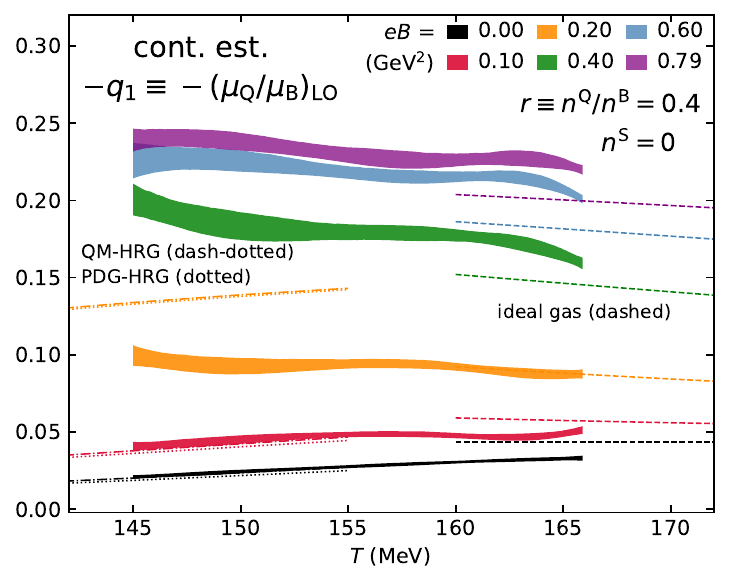}

\caption{Electric charge over baryon chemical potential leading-order coefficient, $-q_1\equiv -\left(\mu_{\rm Q}/\mu_{\rm B}\right)_{\rm LO}$, for strangeness-neutral system $\hat{n}_{\rm S}=0$ with isospin asymmetry parameter $r\equiv n^{\rm Q}/n^{\rm B}=0.4$. Colored bands represent continuum estimates. Left: $eB$-dependence of $-q_1$ at three fixed temperatures $T=145~{\rm MeV}$ (red), $T=155~{\rm MeV}$ (blue) and $T=165~{\rm MeV}$ (gold). Right: $T$-dependence for several fixed strength $eB$ bands spanning vanishing-, weak- and strong-$eB$ regimes. Colored dash-dotted, dotted, and dashed lines represent QM-HRG, PDG-HRG, and magnetized ideal gas results, respectively. The black dashed line in left panel implies the ideal gas free limit $\sqrt{eB}/T \to \infty$ for $T\to \infty$.}  
\label{fig:q1}
\end{figure*}
%%%%

%%%%%
%% Sec: Strangeness-neutral systems
%%%%%
\section{Results under strangeness neutrality and isospin asymmetry}
\label{sec:initial_cond}

In this section, we first present lattice QCD results for the electric charge to baryon chemical potential ratio $q_1$ (Sec.~\ref{subsec:muQovermuB}), followed by the corresponding strangeness to baryon chemical potential ratio $s_1$ (Sec.~\ref{subsec:muSovermuB}). Finally, we systematically explore the dependence of $q_1$ on the isospin parameter $r$, spanning the range from electrically neutral to fully charged baryonic systems (Sec.~\ref{subsec:isospin_r_q1}). 
Throughout this section and subsequent ones, unless explicitly stated otherwise, all lattice QCD results refer to continuum estimates obtained through lattice QCD calculations performed on lattices with temporal extents $N_\tau=8$ and 12~\footnote{A compilation of the $T$--$eB$ plane data and the interpolation/continuum-extrapolation procedure is provided in the Appendix~\ref{app:data}. Compact analytic parameterizations for $q_1(T,eB)$, $s_1(T,eB)$, and $P_2(T,eB)$ are given in the Appendix~\ref{app:param_eos}.}.

\subsection{Ratio $q_1$ of electric charge to baryon chemical potential}
\label{subsec:muQovermuB}

We begin by presenting lattice continuum estimates for the leading-order coefficient $q_1 \equiv q_1(T, eB) = \left( {\muQ}\big/{\muB} \right)_{\rm LO}$ in \autoref{fig:q1}. The left panel shows the $eB$-dependence of $q_1$ at fixed temperatures, indicated by colored bands. To highlight the interplay between magnetic and thermal effects, we select three representative, equally spaced temperature intervals centered around the pseudo-critical temperature at vanishing magnetic field, $T_{pc}(eB=0)$: low-$T = 145~{\rm MeV}$, near-$T_{pc}(eB=0)$ with $T = 155~{\rm MeV}$, and high-$T = 165~{\rm MeV}$. Conversely, the right panel displays $T$-dependence of $q_1$ at various fixed magnetic field strengths, clearly highlighting the systematical evolution from vanishing ($eB=0$), to relatively weak ($eB=0.1~{\rm GeV}^2$), and then strong-$eB$ ($eB=0.2,~0.4,~0.6,~0.79~{\rm GeV}^2$) regimes. Throughout this section, we focus on conditions relevant to heavy-ion collisions involving $Pb$ and $Au$ by fixing the isospin parameter to $r = 0.4$. In Sec.~\ref{subsec:isospin_r_q1}, we will further explore the impact of varying isospin parameters on our results.

We observe in the left panel of~\autoref{fig:q1} that $q_1$ is negative across the entire $T$–$eB$ parameter space, and therefore is presented with a negative sign.  To contextualize our findings, we first note that, even before introducing magnetic fields, $q_1$ exhibits negative values due to the imposed isospin asymmetry constraint characteristic of heavy-ion collision systems. Specifically, for $r=0.4$, the density of positively charged baryons, particularly protons, is suppressed relative to neutrons. In terms of fugacity, this implies that the contributions from positively charged baryons ($\tQ_R > 0$), represented by $e^{\tQ_R \hat{\mu}_\tQ}$, must be suppressed relative to those from neutral baryons ($\tQ_R = 0$), represented by $e^0$. This fugacity-based suppression requires that the electric charge chemical potential must be negative ($\hat{\mu}_\tQ<0$), thus driving the leading-order coefficient $q_1$ negative.

The introduction of magnetic fields further intensifies the negativity of $q_1$, progressively lowering its value as the field strength increases.  To gain insight into this behavior, we turn to predictions from the HRG model in the presence of magnetic fields, as discussed in Sec.~\ref{sec:hrg_igl}. The temperature-colored dash-dotted and dotted lines in \autoref{fig:q1} correspond respectively to QM-HRG and PDG-HRG model calculations. 
In the regime of the relatively weak $eB$ ($eB\lesssim 0.15~{\rm GeV}^2$) and low $T$ ($T\lesssim 155~{\rm MeV}$), both PDG-HRG and QM-HRG agree reasonably well with the lattice continuum estimates, 
suggesting that the HRG framework captures essential aspects of lattice data in this regime. Within the HRG framework, the magnetic field enhances the degeneracy of the lowest Landau level (LLL), particularly influencing charged baryons (cf.~\autoref{eq:dos_eB}). A noteworthy special case is provided by doubly charged spin-$3/2$ baryons, specifically the $\Delta$ baryons, which predominantly contribute to the baryon-electric charge correlation $\chi_{11}^{\rm BQ}$ \cite{Ding:2023bft,Ding:2025jfz}. The enhanced degeneracy of these doubly charged baryons under strong magnetic fields significantly lowers their energy, thereby favoring their relative population compared to neutral baryons. However, despite this enhancement in proton and doubly charged baryon densities, the strict isospin constraint ($r=0.4$) ensures a maintained imbalance, resulting in an even more negative $\hmuQ$ and consequently, an increase in $-q_1$.

%%%%
\begin{figure*}[!htbp]
\centering
\includegraphics[width=0.45\textwidth]{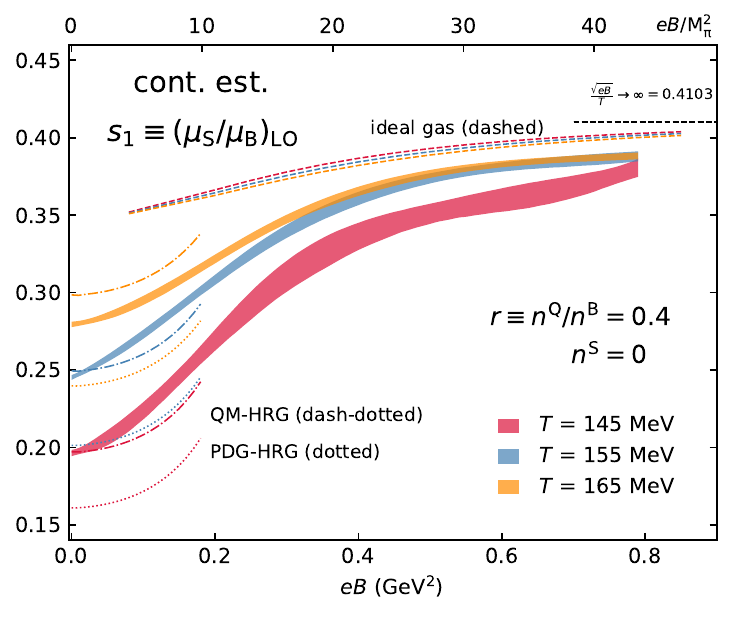}
\includegraphics[width=0.45\textwidth]{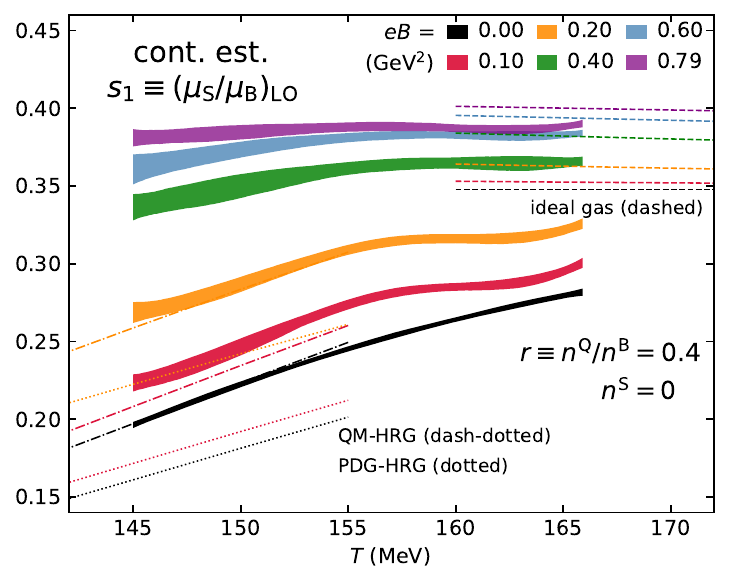}
\caption{Same as~\autoref{fig:q1} but for strangeness over baryon chemical potential leading-order coefficient, $s_1\equiv \left(\mu_{\rm S}/\mu_{\rm B}\right)_{\rm LO}$.
}   
\label{fig:s1}
\end{figure*}
%%%% 

At magnetic fields around $eB \approx 0.15~{\rm GeV}^2$, a noteworthy phenomenon emerges in the left panel of~\autoref{fig:q1}: crossings among fixed-temperature continuum bands become evident, signaling a reversal of the monotonic temperature hierarchy observed at weaker magnetic fields. Such crossings are not captured by HRG models, highlighting their limitation in accounting for nontrivial structural changes arising from intrinsic hadronic structures and non-perturbative QCD effects. Correspondingly, in the right panel of \autoref{fig:q1}, this reversal manifests clearly as a change in the slope of $q_1$ with respect to temperature, transitioning to a monotonic decrease for magnetic fields at $eB=0.2,~0.4,~0.6,~0.79~{\rm GeV}^2$. Additionally, at $eB \simeq 0.15~{\rm GeV^2}$, our lattice QCD results indicate quantitatively distinct enhancement factors for $q_1$ : approximately $3.2$, $2.5$, and $2.0$ at $T = 145$, $155$, and $165~{\rm MeV}$, respectively. This clearly demonstrates a nontrivial interplay between thermal and magnetic effects, primarily driven by the dominant contributions from the highly degenerate lowest Landau level at lower temperatures, reflecting reduced thermal excitation and anticipated dimensional reduction~\cite{Gusynin:1994re}.

At even stronger fields ($eB \sim 0.8~{\rm GeV}^2$), we observe significant enhancement factors of approximately $11$, $8$, and $7$ for temperatures $T$ = 145,~155,~165 MeV, respectively. The reversal of temperature hierarchy in the lattice results qualitatively aligns with the magnetized ideal gas results (see the colored dashed line in~\autoref{fig:q1}), underscoring the dominance of LLL dynamics at strong magnetic fields. Moreover, lattice results progressively approach saturation, converging toward the magnetized ideal gas high-temperature limit, $-q_1(\sqrt{eB}/T \to \infty) = 0.2308$, indicated by the black dashed line in the left panel of \autoref{fig:q1}. Correspondingly, the right panel of \autoref{fig:q1} demonstrates that at these stronger magnetic fields, the gap among fixed magnetic strength bands significantly reduces, reflecting an emergent insensitivity of the electric charge chemical potential to further increases in magnetic field strength. This saturation arises naturally due to the cancellation of the leading linear $eB$ dependence from the dominant LLL in the ratio of generalized susceptibilities defining $q_1$ (cf.~\autoref{eq:q1_def},~\autoref{eq:ideal}). 
%with temperature dependence of $q_1$ also becoming weaker ,

\subsection{Ratio $s_1$ of strangeness to baryon chemical potential}
\label{subsec:muSovermuB}

The effects of the strangeness neutrality condition are directly encoded in the leading-order coefficient 
$s_1\equiv s_1(T,eB)=\left(\muS/\muB\right)_{\text{LO}}$ shown in ~\autoref{fig:s1}. These effects primarily arise through modifications in baryon-strangeness correlations and the electric charge constraint. Analogous to the discussion of $q_1$, the left panel of ~\autoref{fig:s1} shows the magnetic field strength $eB$-dependence of $s_1$ at three fixed temperatures near the QCD transition, while the right panel illustrates its temperature dependence at fixed magnetic field strengths spanning the vanishing-, weak-, and strong-$eB$ regimes. 

Notably, in contrast to its electric charge counterpart $q_1$, $s_1$ is positive due to the negative strangeness quantum number associated with strange quarks. In the absence of magnetic fields, imposing the electric charge constraint arising from isospin asymmetry ($r = 0.4$) skews the strange sector, slightly enhancing $s_1$ to restore strangeness neutrality. Consequently, the Stefan-Boltzmann limit shifts from the naive symmetric $1/3$ limit to $s_1(eB=0, T \to \infty)\approx 0.35$, as noted in~\autoref{tab:free_limit}. With the introduction of magnetic fields, $s_1$ increases, reflecting similar underlying physics to the enhancement of $-q_1$. Specifically, stronger magnetic fields elevate the population of charged strange baryons, analogous to a temperature-induced effect. Consequently, $\hmuS$ must adjust accordingly to maintain strangeness neutrality, resulting in a further rise in $s_1$.

As previously discussed for $q_1$, the HRG model in the presence of magnetic fields provides a valuable framework for interpreting the interplay of thermal and magnetic effects on $s_1$ as well, particularly in regimes of relatively weak $eB$ and low $T$. In ~\autoref{fig:s1}, dash-dotted and dotted colored lines represent the QM-HRG and PDG-HRG results at corresponding fixed temperatures.
Unlike the case for $q_1$, the lattice QCD estimates for $s_1$ differ substantially from PDG-HRG calculations, which include only experimentally observed resonances listed in PDG. In contrast, the QM-HRG calculations, incorporating additional states predicted by the quark model, show significantly improved agreement with lattice data. Thus, this discrepancy for $s_1$ between lattice results and PDG-HRG predictions highlights substantial contributions from additional strange-flavored resonances in the thermal medium, enhancing fluctuations and modifying their correlations with other quantum numbers.

Another prominent feature observed in the left panel of ~\autoref{fig:s1} is the narrowing of gaps between the three fixed-temperature lattice QCD estimate bands with increasing magnetic field strength. Comparing the relative enhancements of $s_1$ at $eB \simeq 0.15~{\rm GeV}^2$ for $T = 145$, $155$, and $165~{\rm MeV}$, we find increases of approximately $25\%$, $18\%$, and $10\%$, respectively. In the right panel of~\autoref{fig:s1}, this trend manifests as a decreasing magnitude of the temperature slope of $s_1$. 
 Such behavior underscores the intricate interplay between thermal and magnetic effects: at sufficiently strong magnetic fields, even relatively low temperatures can significantly populate strange baryons due to enhanced Landau level degeneracy, thereby diminishing the overall temperature dependence of $s_1$.

In regions of extremely strong magnetic fields, the lattice QCD results begin converging towards predictions from the magnetized ideal gas model, represented by the dashed colored lines at fixed temperatures in~\autoref{fig:s1}. With increasing field strength, $s_1$ continues to grow similar to $-q_1$, exhibiting slightly reduced enhancements of approximately 1.9, 1.6, and 1.4 at $eB \sim 0.8~{\rm GeV}^2$ for $T = 145$, $155$, and $165~{\rm MeV}$, respectively. However, in contrast to $q_1$, where
crossings and subsequent reversals of the temperature hierarchy are prominent, $s_1$ exhibits a distinct trend. No crossings between fixed-temperature bands are observed; instead, the left panel shows a gradual merging of these bands, while the right panel shows a consistent reduction in the temperature slope for 
fixed magnetic strength bands. This contrast in behavior between $q_1$ and $s_1$ highlights the intricate interplay between magnetic field effects and the mass scales of quarks, offering valuable insights into how different observables reveal varied reorganizations of QCD matter under extreme conditions.
The absence of a universal magnetic-field-induced crossing further emphasizes that the observable-specific quantum number dependencies dictate the detailed response to strong magnetic fields. This theme will be revisited when examining bulk thermodynamics and the equation of state in subsequent sections.

Similar to $q_1$, $s_1$ is a ratio observable constructed from specific combinations of conserved charge susceptibilities. In the magnetized ideal gas limit, $\sqrt{eB}/T \to \infty$ for $T\to \infty$, the leading linear dependence on $eB$ from the lowest Landau level cancels, resulting $s_1$ to approach its magnetized free-limit value of  $0.4103$. This limit slightly deviates from the naive isospin-symmetric limit of $2/5$, and notably exceeds the free-limit value at vanishing magnetic fields, $s_1(eB = 0, T \to \infty) = 0.35$, as listed in Table~\ref{tab:free_limit}.

\subsection{Isospin parameter $r\equiv n^\tQ/n^\tB$ dependence }
\label{subsec:isospin_r_q1}

%%%%
\begin{figure}[!htbp]
\centering

\includegraphics[width=0.48\textwidth]{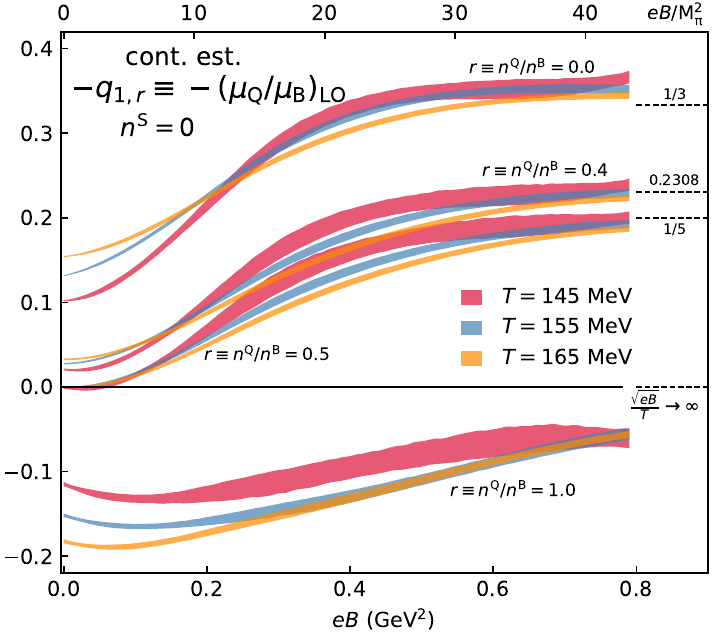}

\includegraphics[width=0.48\textwidth]{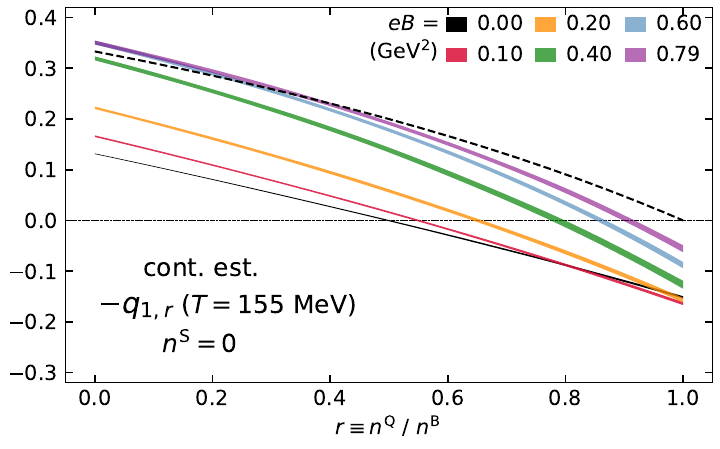}

\caption{Leading-order coefficient of the electric charge to baryon chemical potential ratio, $-q_1 \equiv -\left(\muQ/\muB\right)_{\rm LO}$, for strangeness-neutral systems ($\hat{n}^{\rm S}=0$) with varying isospin parameters. The top panel shows the $eB$-dependence of $-q_1$ for several isospin parameters: $r=0.5$ (symmetric), $r=0.4$ (slightly asymmetric), $r=1.0$ (maximally asymmetric charged), and $r=0.0$ (maximally asymmetric neutral). The bottom panel illustrates the isospin parameter $r$-dependence of $q_1$ at fixed temperature $T=155~{\rm MeV}$. In the top panel, colored bands indicate continuum estimates at $T=145~{\rm MeV}$ (red), $155~{\rm MeV}$ (blue), and $165~{\rm MeV}$ (gold), while in the bottom panel, at fixed magnetic field strengths spanning the vanishing-, weak-, and strong-$eB$ regimes. The black dashed lines mark the ideal gas limits in the presence of magnetic fields  ($\sqrt{eB}/T \to \infty,~T \to \infty$).}
\label{fig:rq1}
\end{figure}
%%%%

In~\autoref{fig:rq1}, we investigate $q_1$ across strangeness-neutral systems characterized by varying isospin parameters $r$. The top panel presents magnetic field dependence of lattice continuum estimates for the isospin-symmetric case ($r = 0.5$), the slightly isospin-asymmetric case relevant to $Pb/Au$ heavy-ion collisions ($r = 0.4$) and the maximally isospin-asymmetric cases---all electrically charged $r=1.0$ and all neutral $r = 0$. Colored bands correspond to three fixed temperatures $T = 145$, $155$, and $165~\mathrm{MeV}$. The bottom panel illustrates the isospin parameter $r$ dependence of $q_1$ at fixed temperature $T=155~{\rm MeV}$, with colored bands indicating continuum estimates for several magnetic field strengths $eB$ spanning vanishing-, weak-, and strong-$eB$ regimes.

In the top panel, for the isospin-symmetric case ($r = 0.5$), we observe that the electrical charge chemical potential, $\hmuQ$, vanishes exactly at $eB = 0$ irrespective of temperature, leading to vanishing $q_{1}(eB = 0) = 0$. This occurs because (2+1)-flavor QCD possesses isospin symmetry due to degenerate light quark masses ($m_u = m_d$), leading to specific symmetry relations among susceptibilities: 
\begin{equation}
2\chi^{\mathrm{QS}}_{11} - \chi^{\mathrm{BS}}_{11} = \chi^{\mathrm{S}}_2, \qquad
2\chi^{\mathrm{BQ}}_{11} - \chi^{\mathrm{BS}}_{11} = \chi^{\mathrm{B}}_2,
\end{equation}
which enforces the exact cancellation of $q_1$. Thermodynamically, isospin symmetry implies equal populations of neutral and charged hadrons, eliminating the need for any suppression or enhancement of charged baryon populations, and hence $\hmuQ = 0$. When isospin asymmetry is introduced, two distinct scenarios arise. For systems with  $r<0.5$, fewer
protons are present relative to neutrons, thus necessitating a negative $\hmuQ$, as previously seen for $r = 0.4$. This effect is even more pronounced for the charge-neutral case ($r = 0$), which yields the largest negative deviation in $q_1$. Conversely, for systems with $r > 0.5$, a positive $\hmuQ$ is required to preferentially populate positively charged baryons and enhance their abundance, 
which is evident in the extreme case $r=1$, representing a system composed entirely of positively charged baryons. The bottom panel further reinforces this interpretation: at $T = 155~{\rm MeV}$ in the absence of magnetic fields, $q_{1,r} \leq 0$ for $r \leq 0.5$ and $q_{1,r} \geq 0$ for $r \geq 0.5$, reflecting the direct correlation between isospin asymmetry and the sign of $q_1$.

With the introduction of a magnetic field, isospin symmetry is explicitly broken due to the different electric charges of the up and down quarks. Consequently, the above-mentioned susceptibility constraints no longer hold (see Ref.~\cite{Ding:2025jfz} for a detailed discussion of these constraints and the extent of isospin symmetry breaking). As established earlier in this section through HRG-based interpretations, magnetic fields enhance the population of positively charged baryons by increasing the degeneracy of their LLL, thus favoring their states over neutral counterparts. To maintain the electric charge constraint in the presence of magnetic fields, $\hmuQ$ has to adjust accordingly. As a result, both top and bottom panels show that $q_1$ decreases as the magnetic field increases for all isospin parameters in strong magnetic fields, reflecting a universal downward trend across the entire $r$-range.

Specifically, for neutron-rich systems ($r<0.5$), where fewer protons are present relative to neutrons, the enhancement of positively charged baryons due to the magnetic field must be compensated by increasingly negative values of $\hmuQ$. Thus $q_1$ becomes progressively more negative. The magnitude of this adjustment strongly depends on both the extent of isospin asymmetry and magnetic field strength. Continuum estimates in both the top and bottom panels consistently illustrate this behaviour: as charge content decreases from $r=0.5 \to 0$ and $eB$ increases, $q_1$ exhibits a progressively stronger negative shift. Remarkably, in the top panel, the difference in $q_1$ among systems characterized by isospin constraints $r\leq0.5$, expressed as $\Delta_r q_{1}(eB,T) = q_{1,r} - q_{1,r = 0}$, appears primarily governed by $r$, exhibiting minimal dependence on magnetic field strength. This trend is echoed in the bottom panel, where magnetic field bands within the $r\leq0.5$ range display approximately uniform slopes, reinforcing the dominance of isospin asymmetry in shaping $q_1$ across this regime.

In contrast, for proton-rich systems ($r > 0.5$), positively charged baryons are naturally favoured, and can result in positive values of $q_1$. As shown in the bottom panel, the transition of $q_1$ to positive values depends sensitively on both the isospin parameter and the magnetic field strength. For the extreme proton-rich case ($r = 1$), composed entirely of positively charged baryons, $q_1$ remains positive across all magnetic field strengths, as seen in both top and bottom panels. However, the magnetic field enhancement of charged baryons necessitates progressively smaller positive adjustments in $\hmuQ$. Consequently, although $q_1$ remains positive, it gradually decreases toward zero with increasing magnetic field strength.

At sufficiently strong magnetic fields, all the systems considered above appear to approach their respective magnetized ideal gas limits, namely $-1/3$, $-0.2308$, $-1/5$, and $0$ for increasing isospin parameter values $r = 0$, $0.4$, $0.5$, and $1.0$, respectively. These asymptotic limits $\sqrt{eB}/T\to \infty$ for $T\to\infty$, taken from~\autoref{tab:free_limit}, are indicated by dashed black lines in~\autoref{fig:rq1}.

%%%%%
%% Sec: Bulk thermodynamics: Pressure
%%%%%

\section{Results for bulk thermodynamics and equation of state}
\label{sec:eos}

%%%%

\begin{figure*}[ht]
\centering

\includegraphics[width=0.45\textwidth]{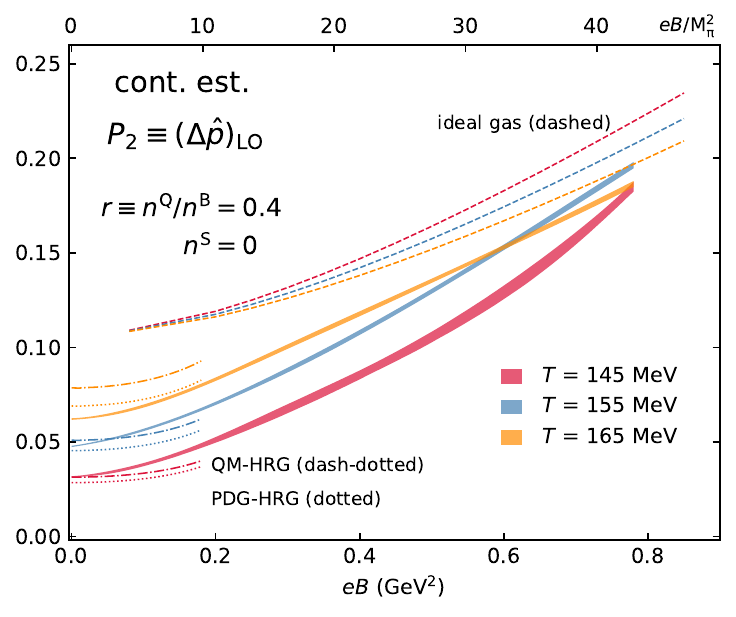}
\includegraphics[width=0.45\textwidth]{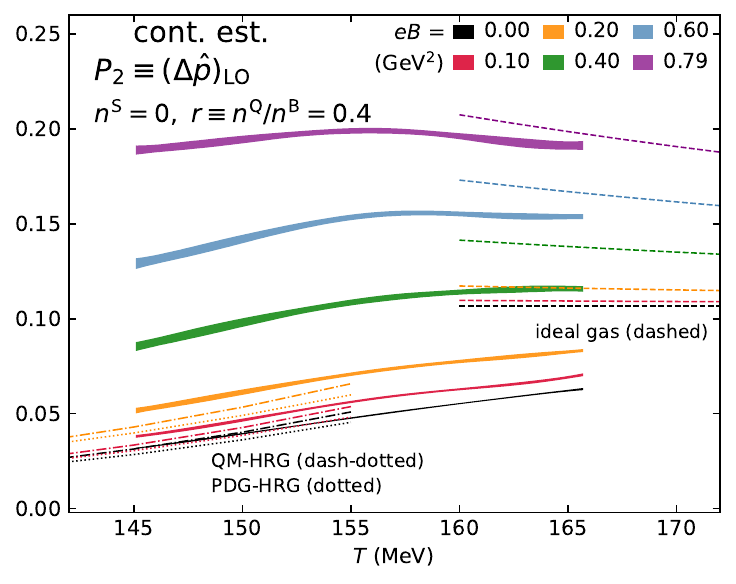}

\caption{Pressure leading-order Taylor expansion coefficient, $P_2\equiv \left(\Delta \hat{p} \right)_{\rm LO}$, for strangeness-neutral system $\hat{n}^{\rm S}=0$ with slight isospin asymmetry $r=0.4$. Colored bands represent continuum estimates. Left: $eB$-dependence of $P_2$ at three fixed temperatures $T=145~{\rm MeV}$ (red), $T=155~{\rm MeV}$ (blue) and $T=165~{\rm MeV}$ (gold). Right: $T$-dependence of $P_2$ for several fixed strength $eB$ bands spanning vanishing-, weak- and strong-$eB$ regimes. Colored dash-dotted, dotted, and dashed lines represent QM-HRG, PDG-HRG, and magnetized ideal gas results, respectively.}   
\label{fig:vseB/p2_Pb_Au_cont}
\end{figure*}
%%%%

In this section, we begin by examining the leading-order coefficient for the pressure coefficient $P_2$ under strangeness-neutral conditions relevant for heavy-ion collisions (Sec.~\ref{subsec:pressure}). We then compare the thermodynamic pressure responses across different collision systems characterized by varying isospin parameters (Sec.~\ref{subsec:pressure_rdep}). Subsequently, we investigate the baryon number density and its ratio with pressure, further elucidating magnetic-field-driven effects (Sec.~\ref{subsec:NPratio}). Lastly, we present lattice QCD results for higher-order bulk observables—trace anomaly, energy density, and entropy density—to provide deeper insight into the structural changes in the QCD equation of state under strong magnetic fields (Sec.~\ref{subsec:energy-like}).

\subsection{Pressure in strangeness-neutral system}
\label{subsec:pressure}

For strangeness-neutral systems, the leading-order pressure coefficient $P_2 \equiv P_2(T,eB) = (\Delta\hat{p})_{\rm LO}$, as introduced in~\autoref{eq:p2k}, can be expressed explicitly in terms of leading-order susceptibilities and chemical potential ratios $q_1$ and $s_1$ as follows
% \bea
% \label{eq:p2_q1s1xxx}
% P_2 &=& \frac{1}{2}\left( \chi^{\tB}_{2}  + \chi^{\tQ}_{2} q_1^2 +  \chi^{\tS}_{2} s_1^2 \right) \nn \\
% && \qquad + ~\chi^{\tB \tQ}_{11} q_1 + \chi^{\tB \tS}_{11}  s_1+ \chi^{\tQ \tS}_{11} q_1 s_1, 
% \eea 
\begin{align}
\label{eq:p2_q1s1}
P_2  &= \frac{1}{2\hmuB} \hat{n}^{\tB}_{\rm LO} (1+rq_1)   \notag \\ 
     &= \frac{ \chi^{\tB}_{2}  + \chi^{\tQ}_{2} q_1^2 +  \chi^{\tS}_{2} s_1^2 }{2}~ + \chi^{\tB \tQ}_{11} q_1 + \chi^{\tB \tS}_{11}  s_1+ \chi^{\tQ \tS}_{11} q_1 s_1\,,
\end{align}
which follows directly from~\autoref{eq:plo} and the strangeness neutrality constraint described in Sec.~\ref{sec:ns0-sys}. The above expression suggests that $P_2$ is significantly driven by strong magnetic field-induced modifications to $q_1$ and $s_1$, as discussed in the previous section.

In \autoref{fig:vseB/p2_Pb_Au_cont}, we present lattice QCD continuum estimates for the leading-order pressure coefficient, $P_2$. The left panel shows $P_2$ as a function of magnetic field strength $eB$ at three fixed temperatures around the QCD transition region, while the right panel displays its temperature dependence at several fixed magnetic field strengths, ranging from vanishing to extremely strong fields, up to $eB = 0.79~{\rm GeV}^2$. These results correspond to strangeness-neutral systems with slight isospin asymmetry ($r=0.4$), conditions relevant for $Pb/Au$ heavy-ion collisions. Later, in Sec.~\ref{subsec:pressure_rdep}, we compare these findings with other collision systems characterized by different isospin parameters and the simplified scenario of vanishing electric charge and strangeness chemical potentials.

We first discuss the lattice QCD results for pressure in the absence of magnetic fields to establish a baseline. At vanishing magnetic field, it is well established that the pressure increases monotonically with temperature as a consequence of thermal agitations~\cite{Bazavov:2017dus}. Across the QCD transition, this monotonic rise becomes notably rapid, reflecting a smooth crossover from hadronic matter to a regime dominated by quark and gluon degrees of freedom. Our lattice results reflect this trend clearly, displaying a distinct temperature hierarchy with $T = 145~\rm{MeV}$ (low temperature), $T = 155~\rm{MeV}$ (near $T_{pc}(eB=0)$),  and  $T = 165~\rm{MeV}$ (high temperature) in the left panel of \autoref{fig:vseB/p2_Pb_Au_cont}, and as a corresponding monotonic growth for the $eB=0$ band in the right panel. At low temperatures, the lattice results exhibit excellent agreement with the HRG model, whereas at high temperatures, they gradually move towards the ideal gas limit. Without the constraint of strangeness neutrality, the leading-order pressure coefficient would simply be $P_2=\chi_2^\tB/2$, and thus its free limit would reach the Stefan-Boltzmann expectation of $1/6$. However, imposing strangeness neutrality systematically reduces this limit. An additional, though smaller, reduction arises from the enforced isospin-asymmetric electric charge constraint, resulting in a final ideal gas limit that falls below the naive expectation.

As magnetic fields are introduced, the behaviour of the pressure becomes significantly more intricate due to the interplay between thermal and magnetic effects. The lattice results for the leading-order pressure coefficient $P_2$ show a clear increase with magnetic field strength across all three fixed temperatures in the left panel of~\autoref{fig:vseB/p2_Pb_Au_cont}. For instance, at $eB = 0.15~{\rm GeV}^2$, $P_2$ is enhanced by approximately $40\%$, $32\%$, and $22\%$ for $T = 145,~155,~165~{\rm MeV}$, respectively.
As highlighted in the previous section for $q_1$ and $s_1$, the HRG framework can provide a reasonable description of lattice results for $P_2$ under relatively weak magnetic fields and low temperatures. Specifically, as detailed in Sec.~\ref{sec:hrg_igl}, the presence of a magnetic field reorganizes the distribution of transverse degrees of freedom for charged particles. The previously continuous momentum states become discretely quantized into Landau levels, whose degeneracy increases proportionally with magnetic field strength. Consequently, the denser distribution of quantum states facilitates greater thermal occupation across temperature ranges, thereby driving an enhancement in pressure with increasing magnetic field strength.

Examining the distinct temperature bands across the QCD transition in more detail in the left panel of \autoref{fig:vseB/p2_Pb_Au_cont}, we observe that the highest-temperature band ($T = 165~\rm{MeV}$) exhibits a slower growth rate of $P_2$ with increasing $eB$ compared to the lower-temperature bands. At the largest field strength $eB=0.79~{\rm GeV}^2$, the enhancement of $P_2$ diminishes progressively with temperature---from approximately a $6$-fold increase at the lowest temperature ($T=145~{\rm MeV}$), to about 4-fold at intermediate temperature ($T = 155~\rm{MeV}$), and finally to 3-fold at the highest temperature ($T = 165~\rm{MeV}$). This differential enhancement leads to a narrowing gap between temperature bands, eventually resulting in crossings among them: first, the intermediate-temperature ($T=155$ MeV) band surpasses the high-temperature ($T=165$ MeV) band at $eB\sim 0.6~{\rm GeV}^2$, followed by the low-temperature ($T=145$ MeV) band crossing the high-temperature band at even stronger fields ($eB \simeq 0.79~{\rm GeV}^2$).
Such crossings clearly indicate a reordering of the temperature hierarchy, marking a qualitative departure from the monotonic hierarchy characteristic of the vanishing- and weak-$eB$ regimes. At these strong magnetic fields, the LLL becomes progressively dominant, leading to pressure enhancements governed primarily by Landau-level degeneracy rather than thermal excitation.  This dominance of the LLL implies a dimensional reduction in the dynamics of charged particles, transitioning from $(3+1)$ to $(1+1)$ dimensions \cite{Gusynin:1994re,Gusynin:1995nb}. 
Consequently, transverse dynamics is effectively quenched due to Landau quantization, while pressure parallel to the magnetic field direction is amplified.
In the regime of extremely strong magnetic fields, lattice results progressively align with magnetized ideal gas predictions, represented by dashed colored lines in \autoref{fig:vseB/p2_Pb_Au_cont}, wherein the temperature hierarchy is ultimately reversed. This reflects the dominance of the LLL and the associated dimensional reduction as the primary mechanisms governing the pressure response under these extreme conditions.

In contrast to the behavior observed for $q_1$ and $s_1$, $P_2$ does not exhibit clear signs of saturation at strong magnetic fields. Within the ideal gas framework, the dominant linear $eB-$dependence originating from the LLL contributions does not cancel out in $P_2$, leading to progressive growth without plateauing, as seen in the left panel of~\autoref{fig:vseB/p2_Pb_Au_cont}. 
Physically, this reflects the extensive nature of pressure, which increases as additional quantum states become accessible due to enhanced magnetic degeneracy. Collectively, the nontrivial crossings among fixed-temperature bands, the emergence of LLL dominance, and the resulting reordering of the temperature hierarchy indicate a fundamental shift in how pressure depends on temperature under extremely strong magnetic fields.

%%%%

\begin{figure}[htbp]
\centering

\includegraphics[width=0.23\textwidth]{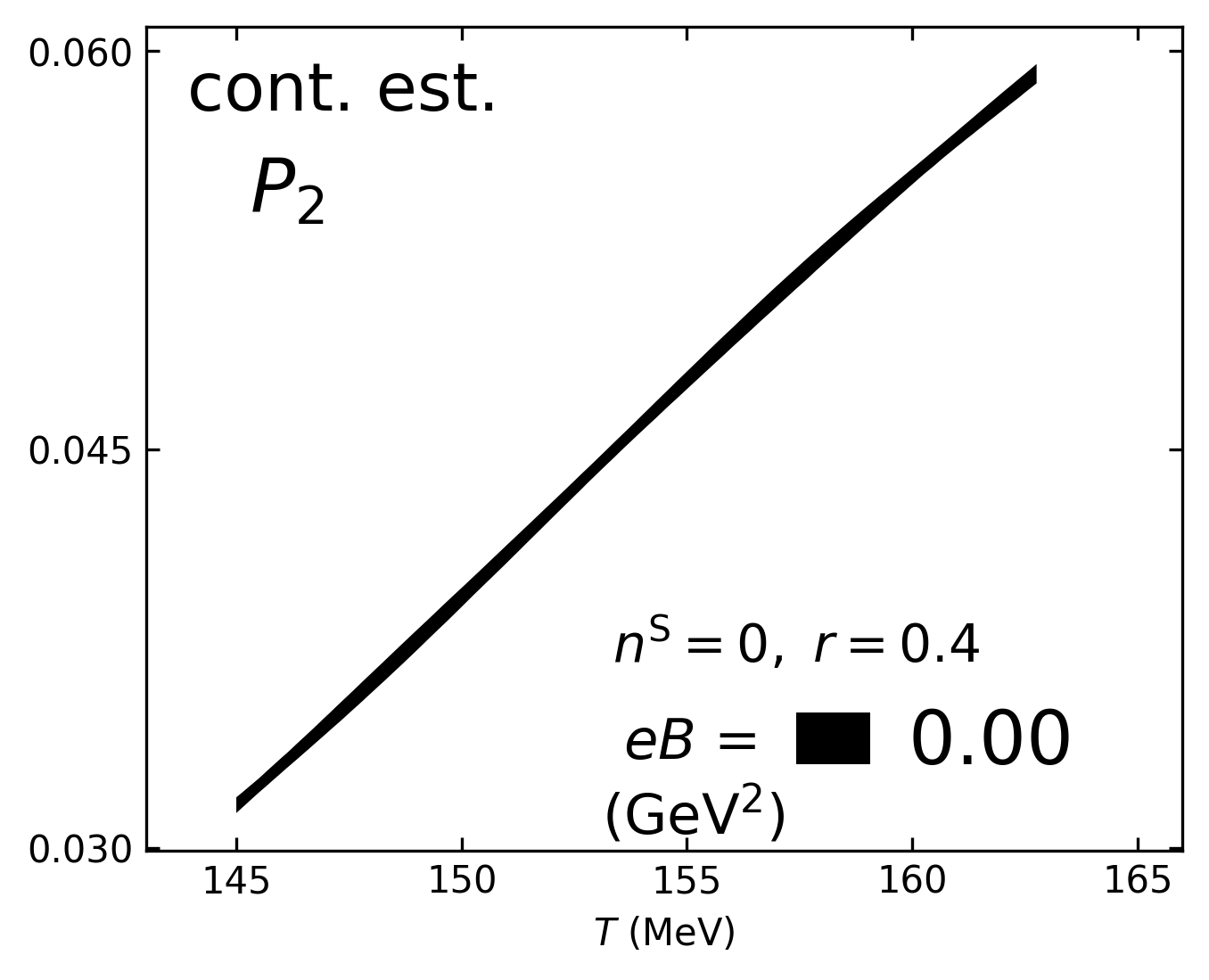}
\includegraphics[width=0.23\textwidth]{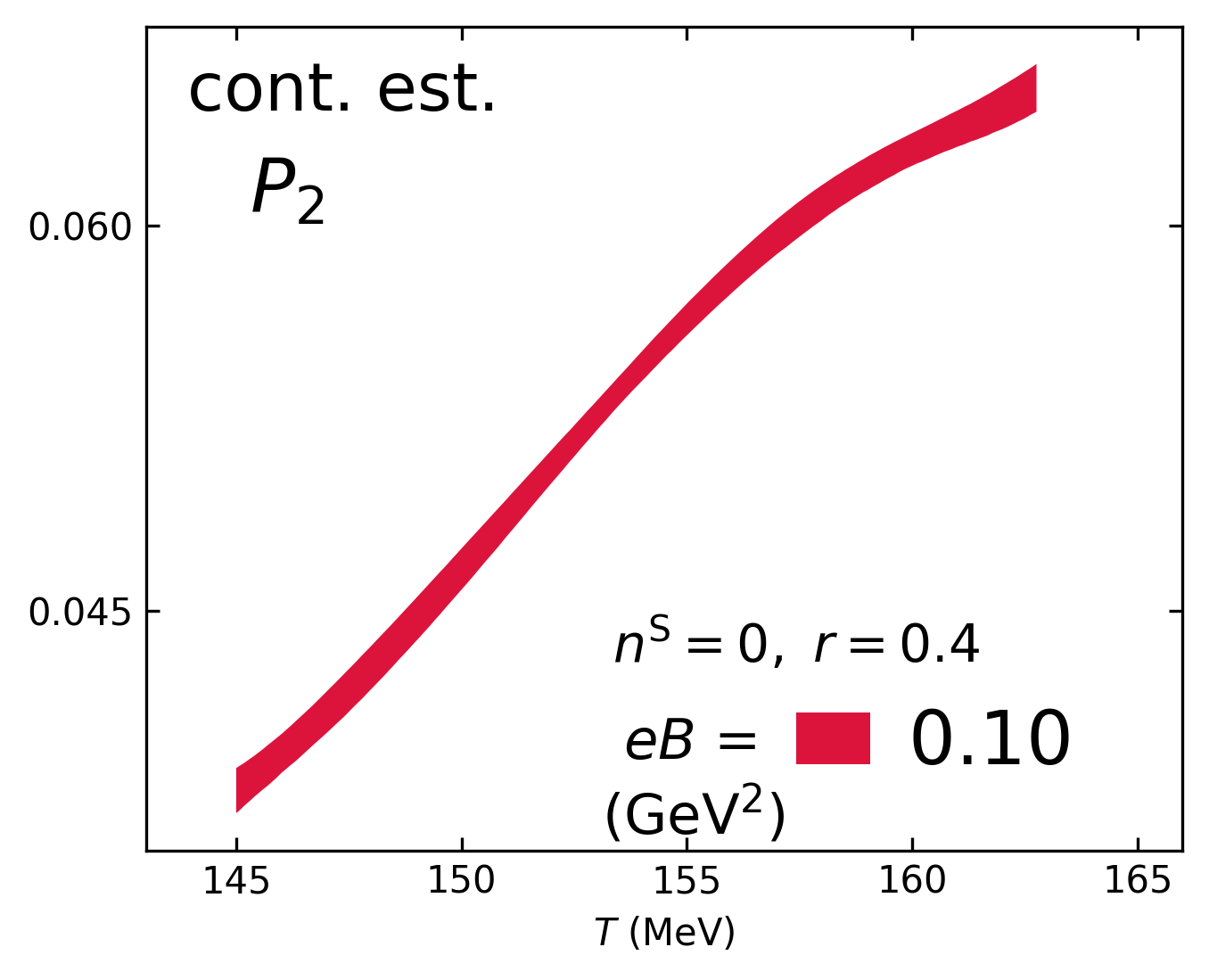}
\includegraphics[width=0.23\textwidth]{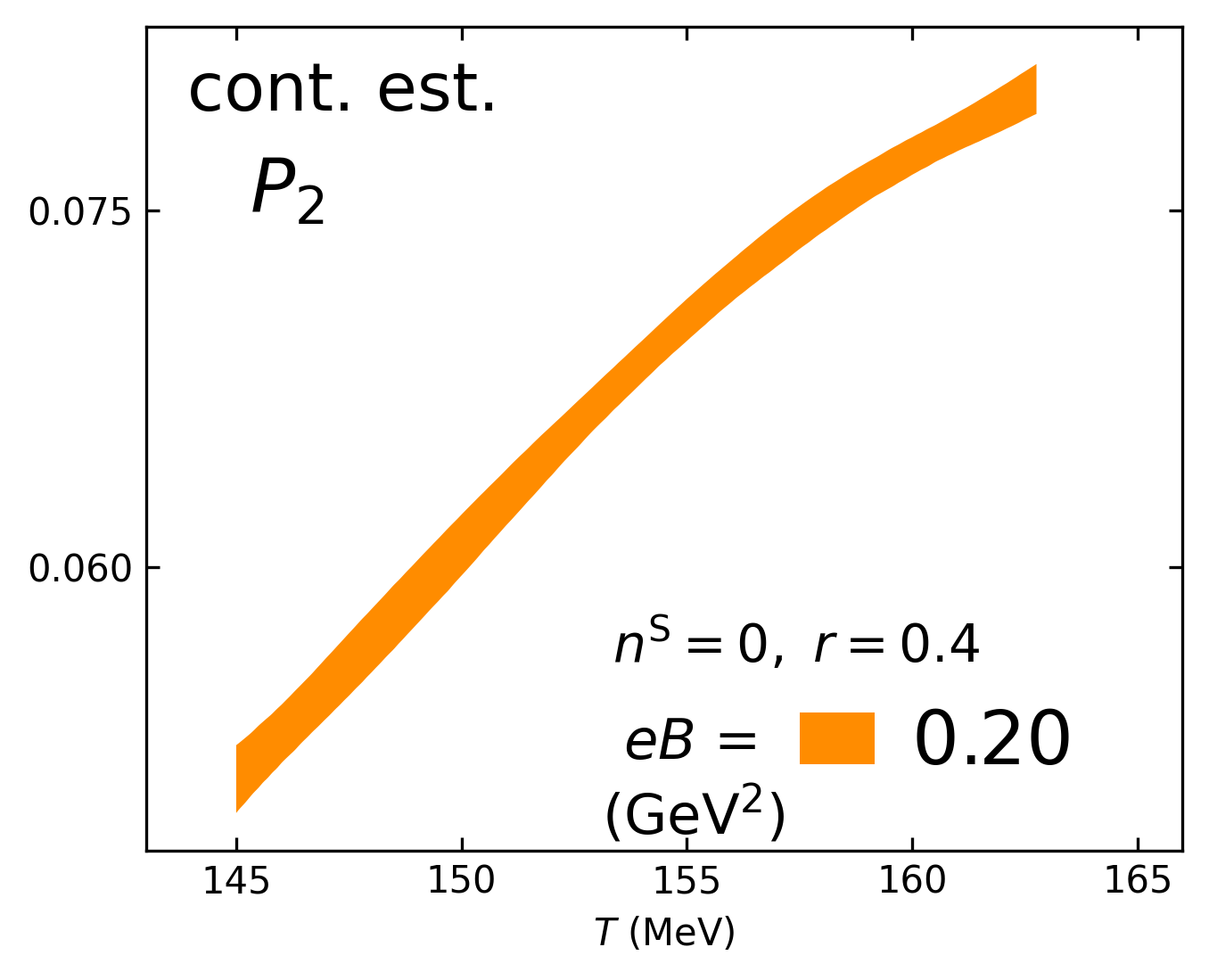}
\includegraphics[width=0.23\textwidth]{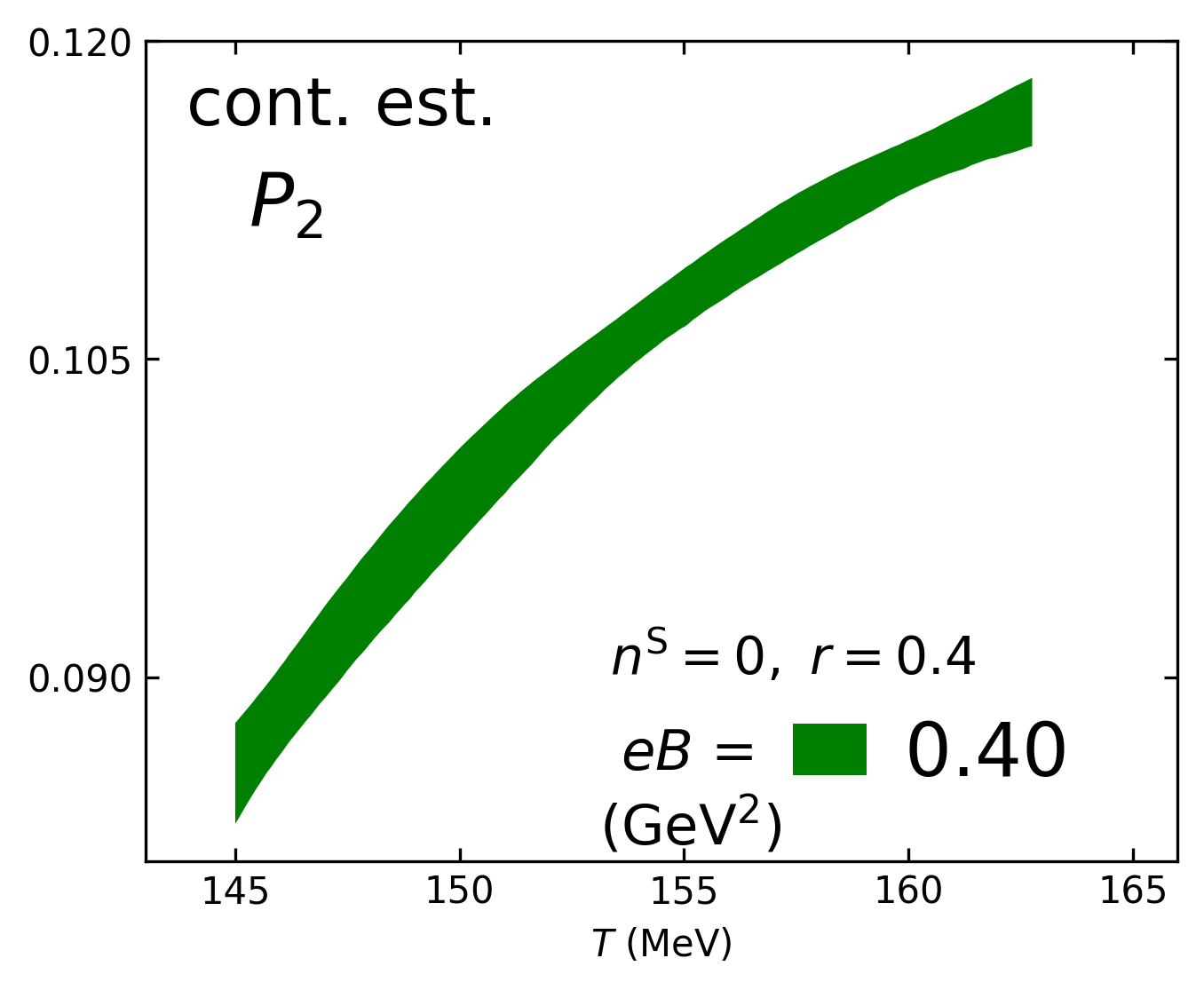}
\includegraphics[width=0.23\textwidth]{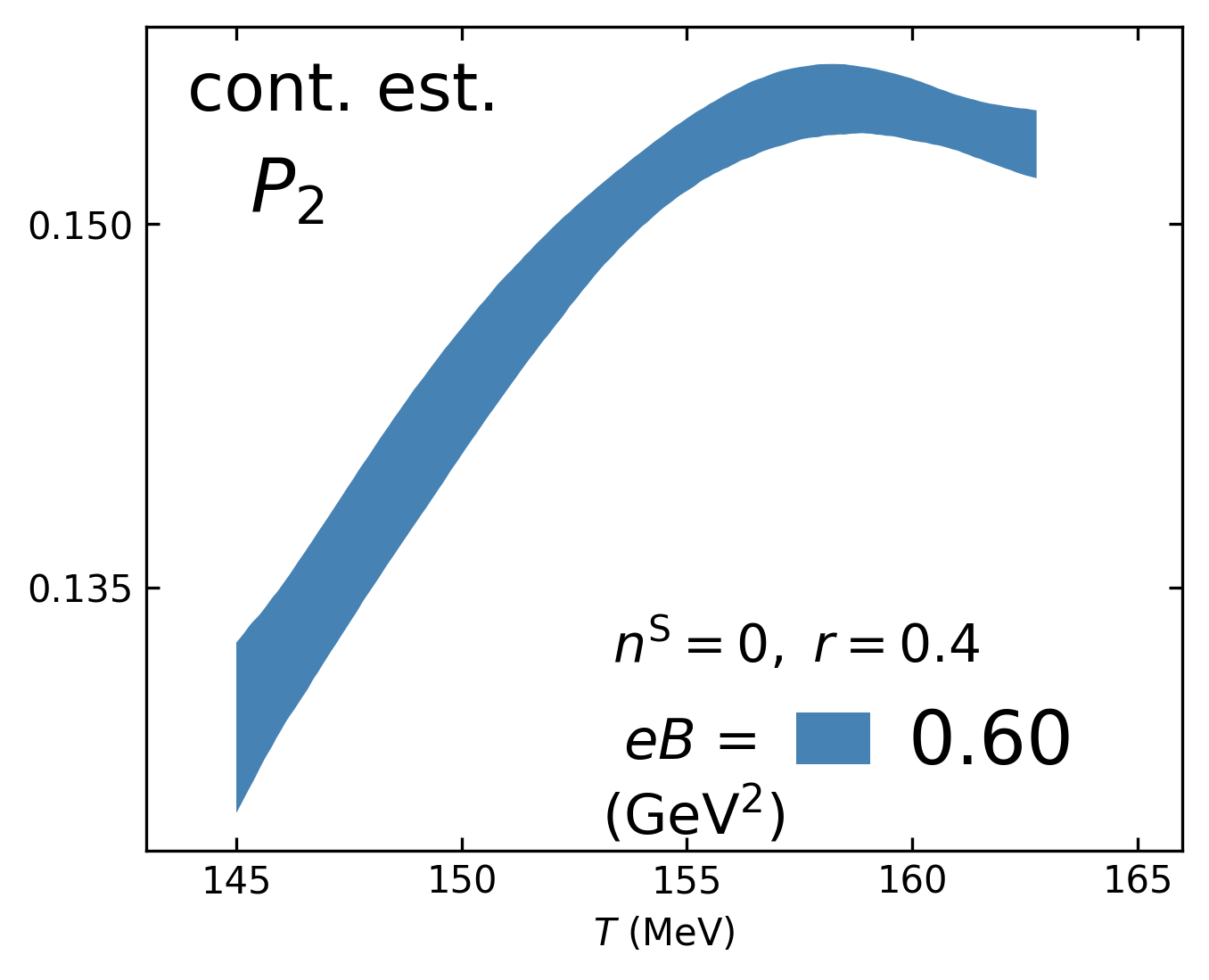}
\includegraphics[width=0.23\textwidth]{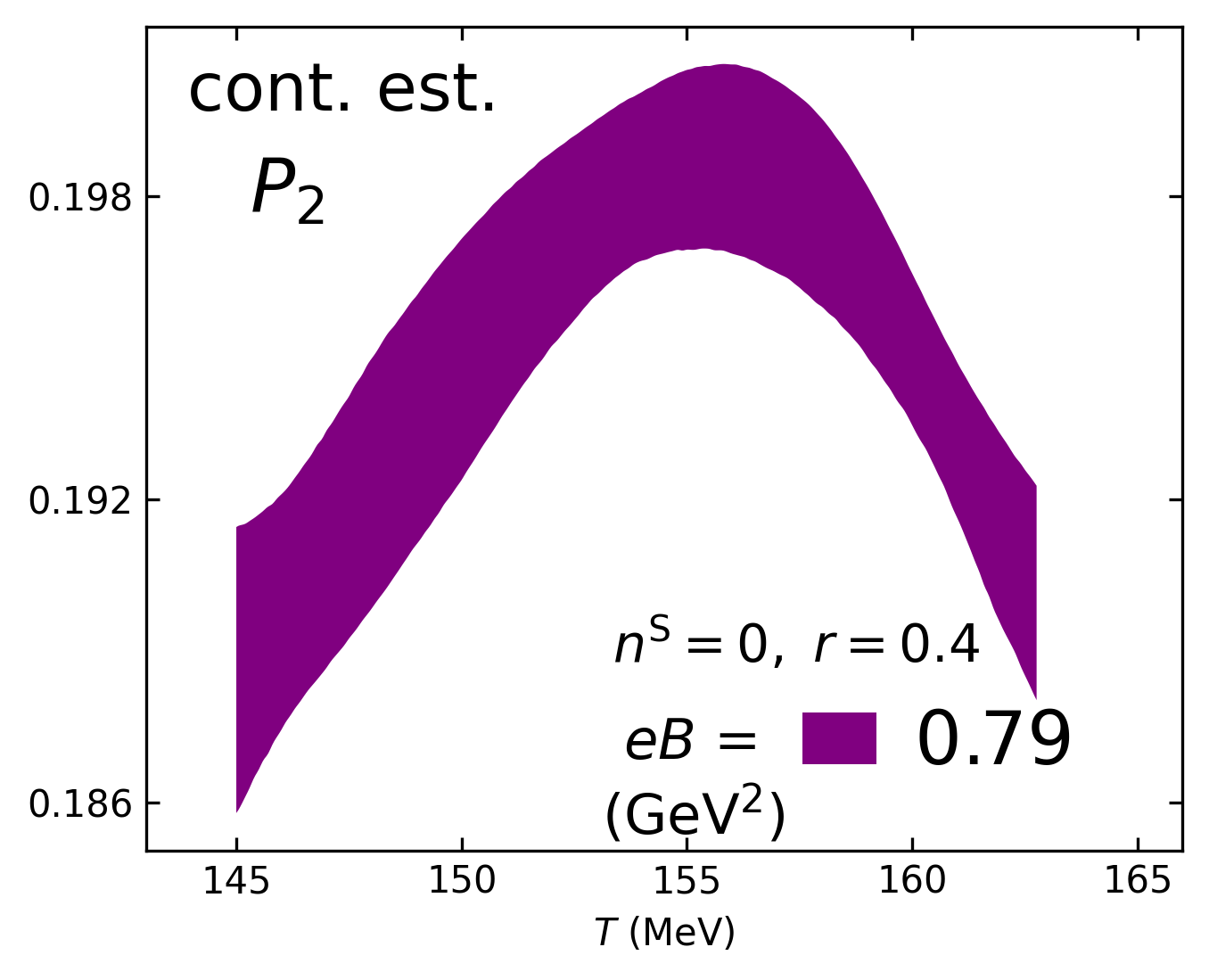}

\caption{Isolated view of the right panel of~\autoref{fig:vseB/p2_Pb_Au_cont} at fixed magnetic field strengths of $eB=0$, 0.1,~0.2~0.4,~0.6~ and 0.79$~{\rm GeV}^2$ from top left to bottom right, showing $T$-dependence of leading-order Taylor expansion coefficients of pressure difference $P_2\equiv \left(\Delta \hat{p} \right)_{\rm LO}$ with $\hat{n}_{\rm S}=0$ and $r=0.4$. }  
\label{fig:vsT/isolate}
\end{figure}
%%%%

Building on the preceding strong magnetic field regime discussion in the left panel, we now examine the corresponding impact on the temperature dependence of $P_2$ in the right panel of Fig.~\ref{fig:vseB/p2_Pb_Au_cont}. As anticipated, increasing the magnetic field strength leads to both rapid quantitative enhancements and significant qualitative modifications in the temperature dependence. Most strikingly, for $eB \simeq 0.6~{\rm GeV}^2$ and beyond--- precisely where crossing among fixed-temperature bands were observed---distinct non-monotonic features begin to develop.  This reveals a direct correspondence between magnetic field-induced reordering and the emergence of non-monotonic structures in the temperature profile. At even stronger magnetic fields, especially at $eB = 0.79~{\rm GeV}^2$, a mild peak structure becomes evident. This behaviour is further illustrated in Fig.~\ref {fig:vsT/isolate}, where the temperature dependence of lattice QCD estimates for $P_2$ is shown for isolated magnetic field strengths, clearly revealing the onset of non-monotonicity for $eB \gtrsim 0.6~{\rm GeV}^2$ and the formation of peak structure at $eB = 0.79~{\rm GeV}^2$.

The inflection point of the leading-order pressure difference coefficient $P_2$ is typically associated with the pseudo-critical temperature $T_{pc}$ of the QCD crossover transition. Notably, our lattice results in Fig.~\ref{fig:vsT/isolate} indicate that both the inflection point and the location of emerging non-monotonic peak structures systematically shift toward lower temperatures with increasing magnetic field strength. This behaviour suggests that the transition in dominant degrees of freedom is driven to occur at reduced temperatures under stronger magnetic fields. In the literature, this downward shift of the transition temperature due to an external magnetic field is commonly referred to as the $T_{pc}$-lowering effect, and is consistent with previous studies based on the light and strange quark chiral condensates~\cite{Bali:2011qj,Ding:2022tqn}, the Polyakov loop~\cite{Bruckmann:2013oba,DElia:2018xwo}, and our recent determination of $T_{pc}(eB)$ using the peak locations of chiral susceptibilities~\cite{Ding:2025jfz}. 

Moreover,  the observed non-monotonic features observed here---namely, the emergence of mild peak structures and the downward shift of inflection points---are consistent with recent equation of state studies at finite lattice spacing $N_\tau = 8$ in $N_f=2+1+1$ QCD~\cite{Borsanyi:2023buy}, as well as investigations at imaginary chemical potential in $N_f=2+1$ QCD~\cite{Astrakhantsev:2024mat}. These findings are further qualitatively supported by our previous studies of conserved charge susceptibilities using heavier-than-physical pion masses~\cite{Ding:2021cwv}, and our latest results obtained at the physical pion mass~\cite{Ding:2025jfz}. Additionally, related studies involving screening masses of pseudoscalar mesons conducted by some of the current authors also support this interpretation~\cite{Ding:2025pbu}.

%%%%
\begin{figure}[!htbp]
\centering

\includegraphics[width=0.48\textwidth]{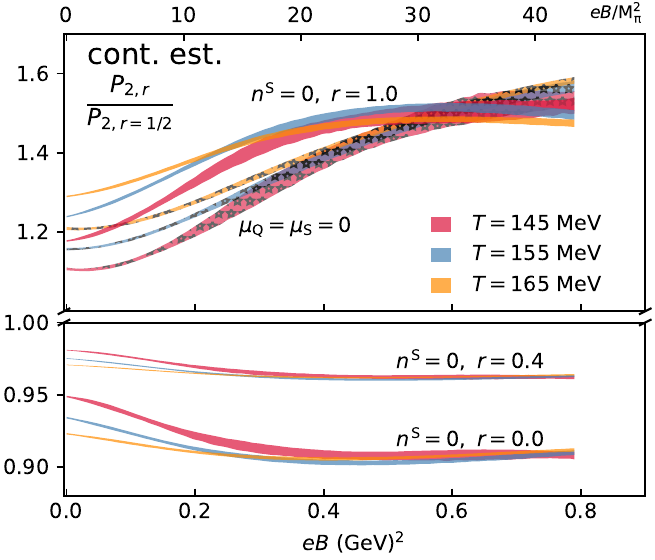}

\includegraphics[width=0.48\textwidth]{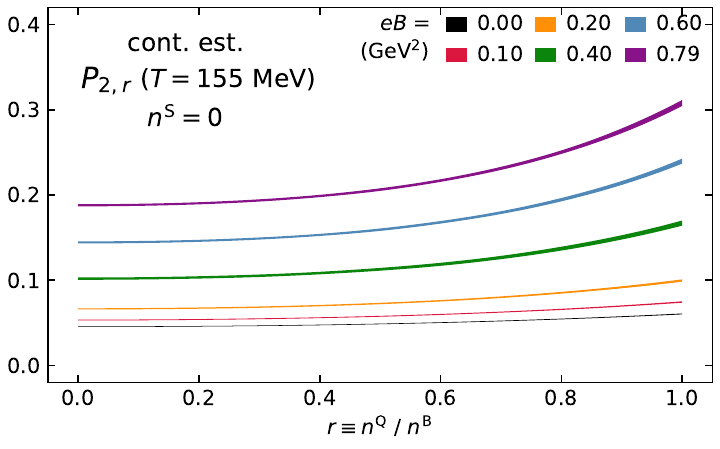}

\caption{Leading-order Taylor expansion coefficient of pressure $P_2$ for various constrained systems. Top: $eB$-dependence of leading order pressure coefficient normalized to its value in the strangeness-neutral and isospin-symmetric system ($\hat{n}_{\rm S}=0,~r=1/2$). Results are shown for the system with vanishing charge and strangeness chemical potentials (hatched) and several strangeness-neutral cases with slight ($r = 0.4$), maximal neutral ($r = 0.0$), and maximal charged ($r = 1.0$) asymmetry. The vertical axis uses distinct scaling above and below unity to enhance visibility. Bottom: Isospin parameter $r$-dependence of $P_{2,r}$ for strangeness-neutral systems at fixed temperature $T=155~{\rm MeV}$. In the top panel, colored bands indicate continuum estimates at $T=145~{\rm MeV}$ (red), $155~{\rm MeV}$ (blue), and $165~{\rm MeV}$ (gold), while in the bottom panel, at fixed magnetic field strengths spanning the vanishing-, weak-, and strong-$eB$ regimes.}

\label{fig:vseB/p2r_155}
\end{figure}

% %%%%

\subsection{Pressure across different constrained systems}
\label{subsec:pressure_rdep}

In general, bulk thermodynamic behaviour is sensitive to constraints imposed on the system, which in turn reshape the baryon potential response of pressure across the conserved charge sectors. Thus far, our discussion of QCD EoS pressure has focused on strangeness-neutral systems with an isospin asymmetry parameter, $r=0.4$, appropriate for $Pb/Au$ heavy-ion collisions. In this subsection, we now explore how varying these constraints influences the thermodynamic pressure response across different collision systems. 

In~\autoref{fig:vseB/p2r_155}, we investigate lattice QCD continuum estimates of $P_2$ for various systems: a system with vanishing electric charge and strangeness chemical potentials ($\hmuQ=\hmuS=0$), and across strangeness-neutral systems characterized by varying isospin parameters $r$. Specifically, the top panel presents magnetic field dependence of $P_2$ normalized with respect to the strangeness-neutral and isospin-symmetric system ($\hat{n}^{\rm S} = 0, ~r = 0.5$), expressed as $P_{2,r}/P_{2,r=1/2}$. Within the strangeness-neutral class, we consider the isospin-symmetric case ($r = 0.5$), the slightly isospin-asymmetric case relevant to $Pb/Au$ heavy-ion collisions ($r = 0.4$), and the maximally isospin-asymmetric cases---all electrically charged $r=1.0$ and all neutral $r = 0$ evaluated for at three fixed temperatures $T=145,~155,~165 ~{\rm MeV}$. The bottom panel further illustrates the isospin parameter $r$ dependence of $P_2$ for strangeness-neutral systems at fixed temperature $T=155~{\rm MeV}$, with colored bands denoting continuum estimates for several magnetic strengths spanning $eB$-parameter space.

%%%%
\begin{figure*}[ht]
\centering

\includegraphics[width=0.45\textwidth]{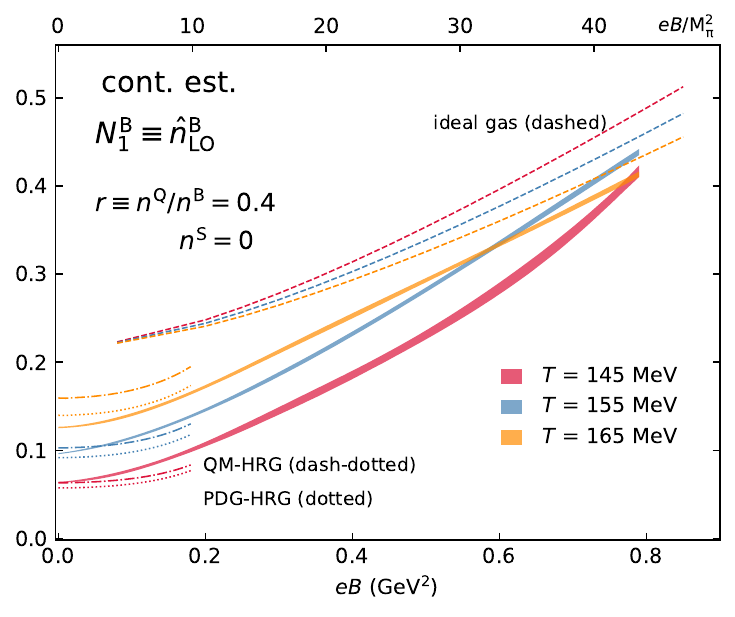}
\includegraphics[width=0.45\textwidth]{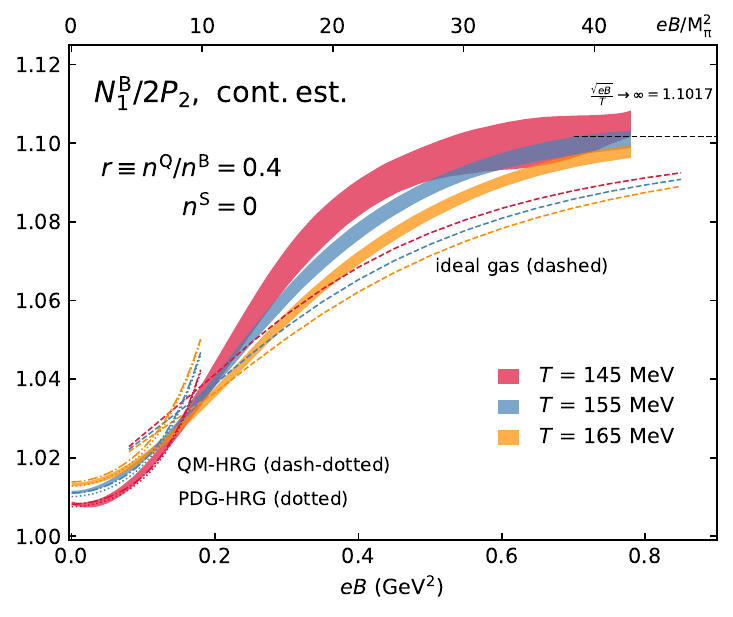}

\caption{Leading-order Taylor expansion coefficients of baryon number density $N^{\rm B}_1\equiv \hat{n}^{\rm B}_{\rm LO}$ (left) and its ratio with pressure $N^{\rm B}_1/2P_2$ (right) versus magnetic field strength $eB$. Results correspond to strangeness-neutral system $\hat{n}^{\rm S}=0$ with slight isospin asymmetry $r=0.4$. Colored bands represent continuum estimates at three fixed temperatures: $T = 145~{\rm MeV}$ (red), $155~{\rm MeV}$ (blue), and $165~{\rm MeV}$ (gold). Colored dash-dotted, dotted, and dashed lines represent QM-HRG, PDG-HRG, and magnetized ideal gas results, respectively. The black dashed line implies the magnetized ideal gas limit $\sqrt{eB}/T \to \infty$ for $T\to \infty$.}  
\label{fig:vseB/nb1_Pb_Au_cont}
\end{figure*}
%%%%

We begin with the simplest case: a system with vanishing electric charge and strangeness chemical potentials ($\hmuQ=\hmuS=0$), represented by hatched bands in the top panel of~\autoref{fig:vseB/p2r_155}. Under these conditions, the leading-order pressure coefficient reduces purely to baryon number fluctuations, $P_2 = \chiB / 2$ according to \autoref{eq:plo}, as no additional constraints restrict the growth driven by the magnetic field-enhanced degeneracy of charged baryons. Consequently, as shown in the top panel of \autoref{fig:vseB/p2r_155}, this scenario exhibits the strongest magnetic response, with the pressure coefficient at $eB \sim 0.8~\text{GeV}^2$ exceeding the strangeness-neutral isospin-symmetric baseline by nearly $60\%$ at $T=145~\text{MeV}$ and continues rising. This unconstrained density scenario thus serves as an upper reference for quantifying the impact of imposed neutrality constraints.

In contrast, strangeness-neutral systems ($\hat{n}^{\rm S}=0$) require dynamically adjusted auxiliary 
chemical potentials $\hmuQ=q_1\hmuB$ and $\hmuS=s_1\hmuB$ to enforce the constraints. Rewriting \autoref{eq:p2_q1s1}, the ratio of pressure coefficients to the isospin-symmetric strangeness-neutral system $(\hat{n}^{\rm S}=0, r=0.5)$ becomes
\begin{equation}
\frac{P_{2,r}}{P_{2,r=1/2}}=\frac{1+rq_{1,r}}{1+0.5q_{1,r=1/2}}\frac{ \hat{n}^{\tB}_{\rm LO,r}}{\hat{n}^{\tB}_{\rm LO,r=1/2}}\,.
\end{equation}

As the isospin ratio decreases from proton-rich ($r=1$) toward neutron-rich ($r=0$) conditions, at both vanishing and nonzero magnetic fields, the coefficient $q_1$ decreases from positive values, crosses zero at $r\geq0.5$ depending on the strength $eB$, and becomes negative, as noted in Sec.~\ref{subsec:isospin_r_q1}. At $eB=0$, the induced electric chemical potential $\hat{\mu}_{\tQ}=q_1\hat{\mu}_{\tB}$ accordingly changes sign, selectively suppressing positively charged baryons for $r<0.5$ and negatively charged ones for $r>0.5$. In the presence of nonzero magnetic fields, although Landau degeneracy enhances available baryonic states, occupancy reductions due to shifts in $\hat{\mu}_{\tQ}$ intensify with increasing $eB$. Consequently, as shown in the top panel of~\autoref{fig:vseB/p2r_155}, for $r>1/2$, the ratio $P_{2,r}/P_{2,r=1/2}$ exceeds unity and grows systematically with stronger fields, reaching enhancements up to $\sim40\%$ at large $eB$. Conversely, for $r<1/2$, this ratio remains below unity and exhibits modestly stronger suppression as $eB$ increases, with deviations reaching approximately $10\%$ at the electrically neutral ($r=0$) scenario. We further look into the $r$ dependence of $P_2$ in the bottom panel of~\autoref{fig:vseB/p2r_155}. In general, $P_{2,r}$ rises monotonically as $r$ increases. Note that, at $r\lesssim0.5$ the $r$ dependence of $P_2$ is mild, and as $r$ increases beyond about 0.5, the dependence becomes stronger and is further intensified at stronger magnetic fields. In particular, at $r=0$, the $eB$-dependence of $P_2$ is dominated by the $q_1\,\chi^{\mathrm{BQ}}_{11}$ term in \autoref{eq:p2_q1s1}(see also Eqs.~\ref{eq:NB1} and \ref{eq:ratio_NB1_P2} below): $q_1$ becomes more negative whereas $\chi^{\mathrm{BQ}}_{11}$ increases as $eB$ grows~\cite{Ding:2023bft}, and their product therefore suppresses the net $eB$-driven rise, leaving $P_2$ with the most muted enhancement in $eB$.

The significant contrast of strangeness neutral systems to those with $\hmuQ=\hmuS=0$ underscores how strangeness neutrality substantially moderates the thermodynamic pressure response under strong magnetic fields. Notably, whether strangeness neutrality is imposed has a far greater impact than specific values of the electric charge-to-baryon ratio $r$ when $r<0.5$. Therefore, accurate modeling of magnetized QCD matter under realistic conditions, such as in heavy-ion collisions ($r\simeq0.4$) and neutron star environments ($r\sim0$), must explicitly incorporate self-consistent treatment of the strangeness sector to avoid systematic overestimation of pressure responses at strong magnetic fields.

\subsection{Baryon number density and ratio with pressure}
\label{subsec:NPratio}

Another fundamental thermodynamic quantity is the conserved charge densities, as defined in~\autoref{eq:nBQS} of Sec.~\ref{sec:CCTE}. The leading-order coefficient for the baryon number density of $\hmuB$ series expansion in~\autoref{eq:nB_muB}, $N^{\rm B}_{1}$, can be explicitly written using~\autoref{eq:nB_LO} as
\bea
N^{\rm B}_1 &=& \chi^{\tB}_{2} + q_1\chi^{\tB \tQ}_{11} + s_1\chi^{\tB \tS}_{11},
\label{eq:NB1}
\eea
which is directly related to the leading-order pressure coefficient $P_2$ through the isospin asymmetry parameter $q_1$, that is,
\beq
{N^{\rm B}_{1} } = {2P_{2}} \left(\frac{1}{1 + rq_{1}} \right).
\label{eq:ratio_NB1_P2}
\eeq

In the left panel of Fig. \ref{fig:vseB/nb1_Pb_Au_cont}, we present the lattice QCD continuum estimates for $N^{\rm B}_1\equiv N^{\rm B}_1(T,eB)=\hat{n}^\tB_{\rm LO}/\hmuB$ as a function of the magnetic field strength $eB$, evaluated at three different fixed temperatures encompassing QCD transition. These results correspond to a strangeness-neutral system with isospin asymmetry parameter $r=0.4$. As anticipated, qualitative behaviour of $N^{\rm B}_1$ closely mirrors that of the leading-order pressure coefficient $P_2$. Notably, the crossing among fixed temperature bands appears to occur at approximately the same magnetic field strengths as observed for $P_2$, implying that $N^{\rm B}_1$ exhibits similar temperature dependence with non-monotonic behavior accompanied by mild peak structures. This similarity highlights an underlying interplay between the magnetic field and thermodynamic constraints governing both baryon number density and pressure.

In the right panel of \autoref{fig:vseB/nb1_Pb_Au_cont}, we show the ratio of the leading-order baryon number density to pressure, ${N^{\rm B}_{1} }/{2P_{2}} \equiv {N^{\rm B}_{1} }/{2P_{2}} (T,eB) = ({\hat{n}^{\rm B} })_{\rm LO}/ (\Delta\hat{p})_{\rm LO}$ as a function of magnetic field strength.  Results for the strangeness-neutral system at $r=0.4$ are presented for three fixed temperatures, along with comparisons to predictions from the HRG model and the magnetized ideal gas limit. Theoretically, deviations of this ratio from unity quantify the extent of isospin symmetry breaking. At vanishing magnetic field, $eB=0$, this deviation remains below $2\%$, which can also be found in Ref.~\cite{Bazavov:2017dus}. Upon introducing magnetic fields, this deviation systematically increases, exhibiting a behaviour resembling that of $q_1$, as discussed in~\autoref{fig:q1} and explicitly supported by the relation in~\autoref{eq:ratio_NB1_P2}. At extremely strong magnetic fields, the ratio approaches the magnetized ideal gas, and ultimately begins to saturate at the ideal gas limit and $\sqrt{eB}/T\to \infty$. Notably, in the high-temperature limit, we observe about $10\%$ strong magnetic field-associated enhancements, that is, increasing from $N^{\rm B}_1/2P_2(eB=0, T\to \infty)=1.018$ \cite{Bazavov:2017dus} to $N^{\rm B}_1/2P_2(\sqrt{eB}/T\to \infty)=1.102$, as summarized in~\autoref{tab:free_limit}.

\subsection{Trace anomaly, energy density and entropy density}
\label{subsec:energy-like}

To gain deeper insights into the equation of state and the redistribution of degrees of freedom under thermal and magnetic effects, we now examine higher-order thermodynamic energy-like observables---namely trace anomaly, energy density, and entropy density---as defined earlier in Sec. \ref{sec:CCTE}. Starting from their respective expansions in terms of $\hmuB$, as given in Eqs.~\ref{eq:theta2k_muB},~\ref{eq:eps2k_muB}, and~\ref{eq:sig2k_muB}, the leading-order coefficients, $\Theta_2\equiv  \Theta_2(T,eB)$, $\epsilon_2\equiv  \epsilon_2(T,eB)$, and $\sigma_2\equiv  \sigma_2(T,eB)$, can be expressed explicitly through the leading-order thermodynamic coefficient introduced thus far, that is, $P_2, {N^{\rm B}_{1} },q_1$ and $s_1$, as well as their temperature derivatives:
\bea
\Theta_{2} &=& -rT~\frac{\partial q_1}{\partial T}~ N^{\rm B}_1  +T ~\frac{\partial P_2}{\partial T},\\
\epsilon_{2} &=& \Theta_{2} +3P_2,\\
\sigma_{2} &=& \epsilon_{2} +P_2 - (1+rq_1)N^B_1.
\eea
%%%%
\begin{figure}[!t]
\centering

\includegraphics[width=0.42\textwidth]{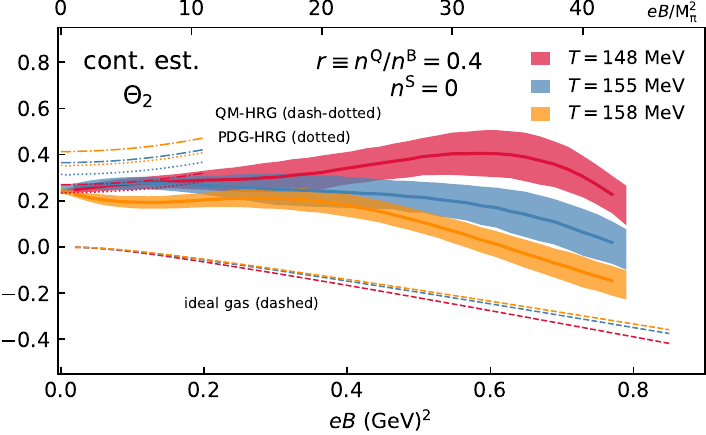}

\includegraphics[width=0.42\textwidth]{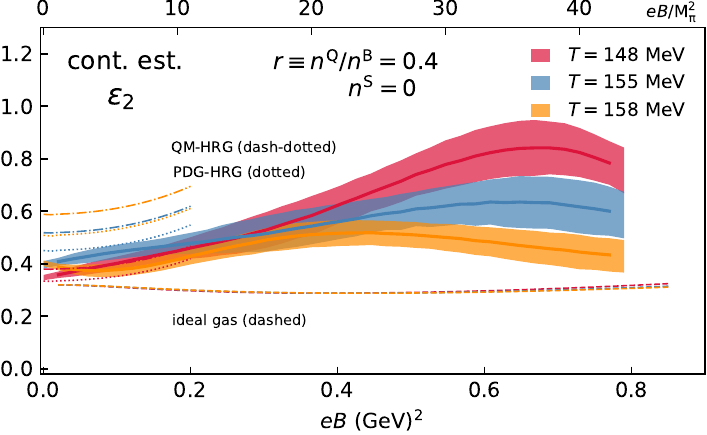}

\includegraphics[width=0.42\textwidth]{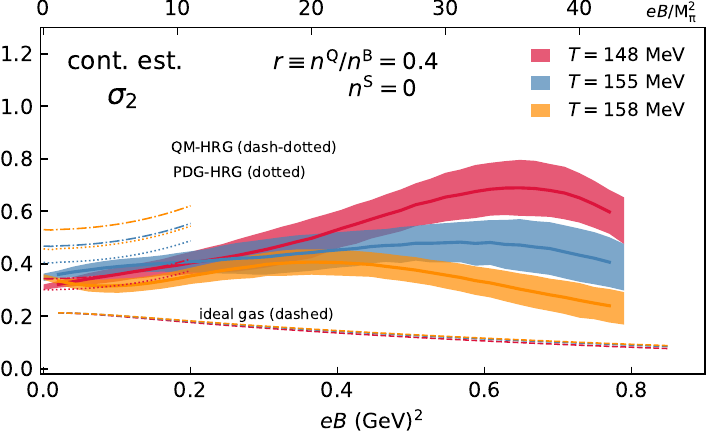}

\caption{Leading-order Taylor expansion coefficients of trace anomaly $\Theta_2$ (top), and energy $\epsilon_2$ (middle), and entropy density $\sigma_2$ (bottom) for strangeness-neutral system $\hat{n}_{\rm S}=0$ with slight isospin asymmetry $r=0.4$.  Colored bands represent continuum estimates versus field strength $eB$ at three fixed temperatures $T=148~{\rm MeV}$ (red), $T=155~{\rm MeV}$ (blue) and $T=158~{\rm MeV}$ (gold). Colored dash-dotted, dotted, and dashed lines represent QM-HRG, PDG-HRG, and magnetized ideal gas results, respectively.}  
\label{fig:vseB/eps2_sig2_theta2_Pb_Au_cont}
\end{figure}
%%%%

\begin{figure*}[!t]
\centering

\includegraphics[width=0.323\textwidth]{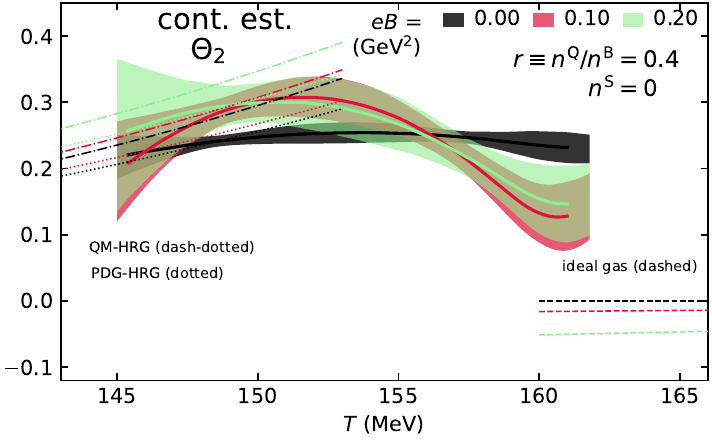}
\includegraphics[width=0.323\textwidth]{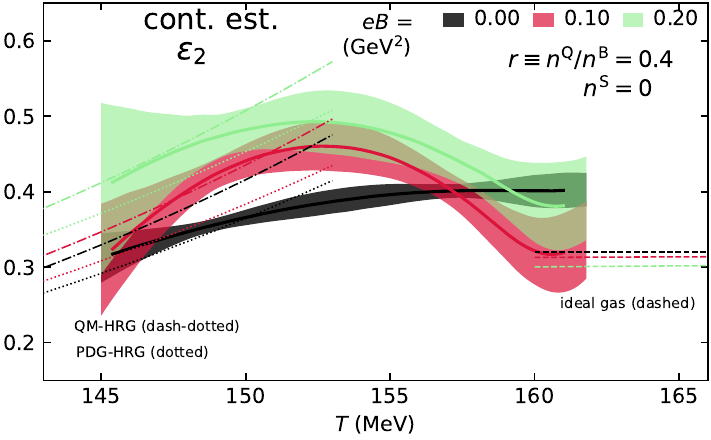}
\includegraphics[width=0.323\textwidth]{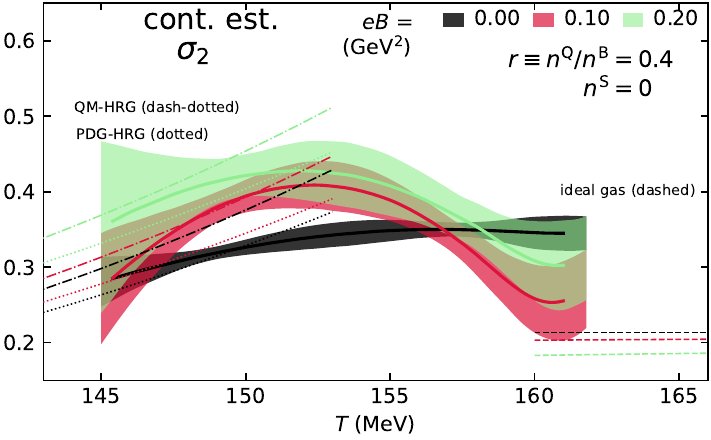}

\includegraphics[width=0.323\textwidth]{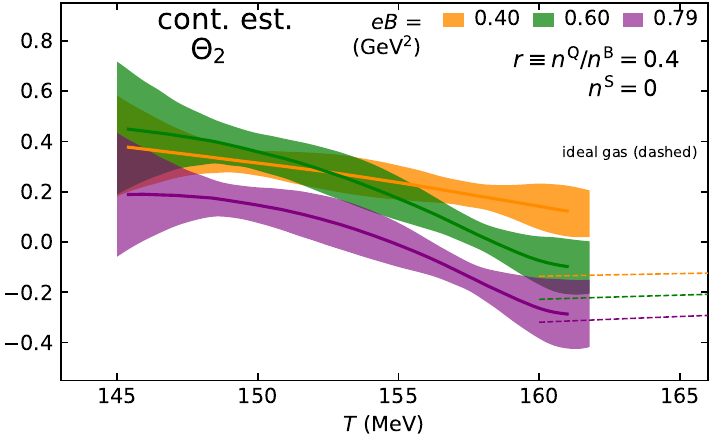}
\includegraphics[width=0.323\textwidth]{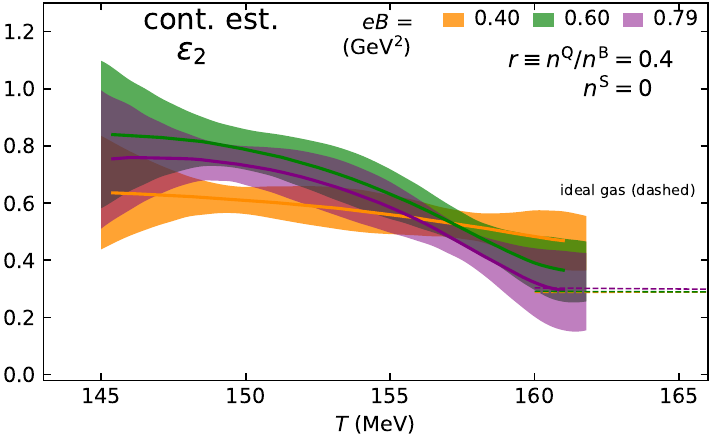}
\includegraphics[width=0.323\textwidth]{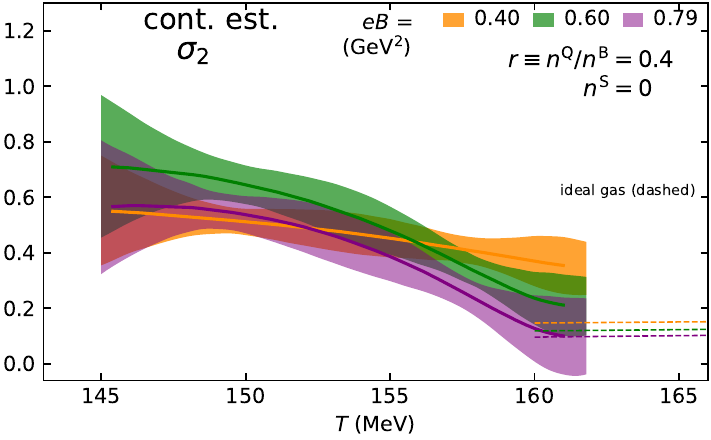}

\caption{Leading-order Taylor expansion coefficients of trace anomaly $\Theta_2$ (left panel), and energy $\epsilon_2$ (middle panel) and entropy density $\sigma_2$ (right panel) for strangeness-neutral system $\hat{n}_{\rm S}=0$ with slight isospin asymmetry $r=0.4$. The top and bottom panels correspond to the vanishing- and weak-$eB$ $(eB=0.0,~0.1,~0.2~{\rm GeV}^2)$, and the strong-$eB$ $(eB=0.4,~0.6,~0.79~{\rm GeV}^2)$ regimes, respectively. Colored bands represent continuum estimates, and colored dash-dotted, dotted, and dashed lines represent respective QM-HRG, PDG-HRG, and magnetized ideal gas results.}  
\label{fig:vsT/eps2_sig2_Pb_Au_cont_vsT}
\end{figure*}

These expressions reveal that the expansion coefficients for energy-like observables exhibit a more intricate dependence on both temperature and magnetic field compared to pressure. In particular, they explicitly incorporate the temperature derivatives of $P_2$ and $q_1$, thus providing critical insight into how magnetic fields modify their thermal behavior near the QCD transition.

In Fig.~\ref{fig:vseB/eps2_sig2_theta2_Pb_Au_cont}, we present lattice results for the leading-order coefficients of the trace anomaly $\Theta_2$ (top), energy density $\epsilon_2$ (middle), and entropy density $\sigma_2$ (bottom) as functions of magnetic field strength $eB$. The calculations are carried out for strangeness-neutral systems with fixed isospin $r=0.4$. The colored bands represent continuum estimates for three fixed temperatures, $T=148~{\rm MeV}$, $T=155~{\rm MeV}$, and $T=158~{\rm MeV}$. These temperatures span the vicinity of the QCD transition region, including the regime where the non-monotonic peak structure of the pressure coefficient $P_2$ occurs under extremely strong fields, as discussed earlier.

In the regimes of vanishing and relatively weak magnetic fields, continuum estimates of $\Theta_2$ at different temperatures appear closely clustered. This clustering reflects the presence of an inflection point in the pressure near the pseudo-critical temperature. Such a behavior is notably absent in 
the QM-HRG and PDG-HRG models, which instead predict a simple monotonic rise with temperature. Similar discrepancies are also evident for $\epsilon_2$ and $\sigma_2$.

As the magnetic field strength increases, notable features emerge in $\Theta_2$. For temperatures slightly below the pseudo-critical temperature $T_{pc}(eB)$~\cite{Ding:2025jfz}, $\Theta_2$ is enhanced, signaling positive deviations from conformality, and exhibits peak structures---for instance, at $T = 148~{\rm MeV}$, it peaks near $eB \sim 0.6~{\rm GeV}^2$. In contrast, for temperatures near or above $T_{pc}(eB)$, $\Theta_2$ becomes progressively suppressed and eventually turns negative. Specifically, at $eB \sim 0.8~{\rm GeV}^2$, where the peak structure in $P_2$ was observed in the bottom right plot of~\autoref{fig:vsT/isolate}, $\Theta_2$ nearly vanishes at $T = 155~{\rm MeV}$, indicating proximity to the maximum of the pressure response. At this field strength, $\Theta_2$ remains slightly positive at $T = 148~{\rm MeV}$ but clearly turns negative at $T = 158~{\rm MeV}$, corresponding respectively to temperatures below and above the peak temperature in $P_2$. 
This significant sign change explicitly highlights a regime where the thermodynamic contribution of the pressure dominates over the energy density, satisfying $3P_2>\epsilon_2$. The negative trace anomaly coefficient thus strongly emphasizes a scenario dominated by pressure-driven dynamics, clearly departing from the naive expectation of near-conformality $(\epsilon_2-3P_2) \to 0$. Such behavior potentially carries profound implications for the hydrodynamic evolution and the thermalization processes in strongly magnetized QCD matter produced in heavy-ion collisions or magnetars.

$\epsilon_2$ and $\sigma_2$ also reflect these non-trivial modifications captured in $\Theta_2$. We observe that continuum estimates of both $\epsilon_2$ and $\sigma_2$, grow with increasing strength $eB$ across all considered temperatures in the weak$-eB$ regime, primarily driven by the enhanced thermodynamic response of pressure. In the strong-$eB$ regime, lower temperature bands exhibit higher energy and entropy densities. Furthermore, different temperature bands appear to peak at distinct magnetic field strengths, followed by a gradual decline as $eB$ increases. For extremely strong fields, the ideal gas limit serves as a useful benchmark, characterized by the dominance of highly degenerate lowest Landau level states and signifying an effective dimensional reduction of the system. In this context, at sufficiently high temperatures and magnetic fields, all three energy-like observables approach their respective ideal gas limits.

In \autoref{fig:vsT/eps2_sig2_Pb_Au_cont_vsT} we further investigate the temperature dependences of $\Theta_2$ (left panel), $\epsilon_2$ (middle panel), and $\sigma_2$ (bottom panel) for a strangeness-neutral system with an isospin asymmetry parameter $r=0.4$. The top panel shows results for vanishing and relatively weak magnetic fields $(eB=0.0,~0.1,~0.2~{\rm GeV}^2)$, while the bottom panel corresponds to the strong-$eB$ $(eB=0.4,~0.6,~0.79~{\rm GeV}^2)$ regime.  The colored bands represent continuum estimates for fixed magnetic field strengths. Magnetized ideal gas results at higher temperatures ($T\gtrsim160~{\rm MeV}$) are included as dashed lines for reference.

In the regime of relatively weak magnetic fields, shown in the top panel of~\autoref{fig:vsT/eps2_sig2_Pb_Au_cont_vsT}, where~\autoref{fig:vseB/eps2_sig2_theta2_Pb_Au_cont} previously showed closely clustered $\Theta_2$ bands at fixed $T$, the $T$-dependence results of all three energy-like observables exhibit shallow maxima slightly below the pseudo-critical temperature $T_{pc}(eB)\simeq T_{pc}(eB=0)$. This mild non-monotonicity is consistent with recent QCD EoS studies without a magnetic field, where peak temperature corresponds to the inflection point of pressure near $T_{pc}(eB=0)$ for isospin-symmetric, strangeness-neutral systems~\cite{Bollweg:2022fqq}. For $T\lesssim155~{\rm MeV}$, the QM-HRG and PDG-HRG curves are also displayed using dash-dotted and dotted lines, respectively. As discussed above, both HRG models fail to capture the non-monotonic behavior observed in the lattice QCD data, underscoring their limitations in describing the thermodynamics of the QCD transition under magnetic fields.

In the strong-$eB$ regime, shown in the bottom panel of~\autoref{fig:vsT/eps2_sig2_Pb_Au_cont_vsT}, the $T$-dependence becomes more intricate due to non-monotonic contributions from pressure. As $T$ increases, all three energy-like coefficients exhibit a systematic decline, indicating that the peak structures observed in the weak-$eB$ regime are possibly shifted to $T<145~\rm MeV$ or obscured by current uncertainties. The characteristic trends seen in the $eB$-dependence in~\autoref{fig:vseB/eps2_sig2_theta2_Pb_Au_cont} are mirrored here: for $eB\lesssim 0.6~{\rm GeV}^2$, the magnitudes of $\Theta_2,\epsilon_2$ and $\sigma_2$ at low temperatures ($T\lesssim T_{pc}(eB)$~\cite{Ding:2025jfz}) increase progressively with magnetic fields strength. Around $eB \sim 0.6~\mathrm{GeV}^2$, the pressure coefficient $P_2$ exhibits non-monotonic behaviour (bottom left of~\autoref{fig:vsT/isolate}), and the associated $\Theta_2$, sensitive to both the temperature derivative of $P_2$ and isospin asymmetry, turns negative at higher $T$. For $eB \gtrsim 0.6~\mathrm{GeV}^2$, the overall magnitudes of all energy-like coefficients decrease with increasing $eB$, as seen by comparing the $eB=0.6~\mathrm{GeV}^2$ to $ eB= 0.79~{\rm GeV}^2$ bands. This indicates that the peak in $P_2$ is followed by a reduction in both $\epsilon_2$ and $\sigma_2$. At $eB=0.79~{\rm GeV}^2$, $\Theta_2$ vanishes at $T\simeq155~{\rm MeV}$---coinciding with a mild peak structure in $P_2$---and changes sign across this temperature. Although the detailed $T$- and $eB$-dependencies may vary among bulk thermodynamic observables, the emergence of such non-monotonic structures at strong magnetic fields points to a similar underlying non-perturbative interplay between thermal and magnetic effects.

\section{Conclusions}
\label{sec:summary}

In this work, we have presented continuum-estimated $(2+1)$-flavor lattice QCD results for the leading-order Taylor expansion coefficients of bulk thermodynamic observables in the presence of strong magnetic fields and nonzero baryon chemical potential. Calculations were carried out using the HISQ action at the physical pion mass, covering magnetic field strengths up to $eB \simeq 0.8~\mathrm{GeV}^2$ and temperatures in the crossover region, $145 \lesssim T \lesssim 165~\mathrm{MeV}$. The primary focus was on strangeness-neutral systems with a slight isospin asymmetry ($r \equiv n_\tQ/n_\tB = 0.4$), relevant for $Pb/Au$ heavy-ion collisions, while also examining the response at other values of $r$.

For the baseline case $r=0.4$, $q_1 \equiv (\muQ/\muB)_{\mathrm{LO}}$ becomes more negative with increasing $eB$ and, in the strong-field regime, exhibits crossings among fixed-temperature bands, leading to a reversal of the temperature ordering. In contrast, $s_1 \equiv (\muS/\muB)_{\mathrm{LO}}$ remains positive across the entire $eB$ range and shows no such crossings. The $r$-dependence of $q_1$ is systematic: lowering $r$ (more neutron-rich) makes $q_1$ more negative at fixed $eB$, so that $|q_1|$ is larger for $r=0$ than for $r=0.4$, while increasing $r$ (more proton-rich) shifts $q_1$ toward positive values.

The leading-order pressure coefficient $P_2$ increases monotonically with $eB$ for all studied temperatures; in the strong-field regime it also develops crossings among fixed-$T$ bands and a reversal of the temperature hierarchy, consistent with the onset of lowest-Landau-level dominance. At $r=0.4$, the leading-order baryon number density coefficient $N_{1}^\tB$ tracks $P_2$ in its $T$--$eB$ trends. The ratio $ N_{1}^\tB /(2P_2)$, which quantifies isospin-symmetry breaking, starts within $\sim2\%$ of unity at $eB=0$, grows monotonically with $eB$, and reaches about $1.10$ by $eB\simeq0.8~\mathrm{GeV}^2$ (depending on $T$), approaching the magnetized free (ideal-gas) limit at very large $\sqrt{eB}/T$. 

Looking beyond the baseline $r=0.4$ system, we find that for $r=0$ (charge-neutral matter), the magnetic amplification of $P_2$ is the most muted among the cases examined, despite $|q_1|$ being larger than in the $r=0.4$ case. This $r=0$ limit provides a useful qualitative reference for charge-neutral, magnetized baryonic matter, relevant to neutron-star studies, noting that our results are for hot, strangeness-neutral matter rather than cold, $\beta$-equilibrated conditions.

The leading-order coefficients of the trace anomaly $\Theta_2$, energy density $\epsilon_2$, and entropy density $\sigma_2$ exhibit a coherent pattern connected to the behavior of $P_2$: enhancement with $eB$ at low $T$, mild non-monotonicity near intermediate fields, and decreasing magnitudes toward the largest $eB$; at high $T$ and strong fields, $\Theta_2$ can vanish or turn negative. These correlated features point to a common non-perturbative origin associated with Landau-level reorganization.

Comparisons with the hadron resonance gas (HRG) model show that, in the low-$T$ region and for relatively weak $eB$, the lattice results for $q_1$, $s_1$, and $P_2$ are broadly consistent with HRG expectations, including the effects of Landau quantization of charged hadrons. However, deviations become visible as $T$ approaches the crossover or as $eB$ increases, where the HRG description gradually breaks down and quark--gluon degrees of freedom become essential.

In summary, our continuum-estimated results demonstrate that strong magnetic fields qualitatively restructure the temperature dependence of leading-order bulk thermodynamic coefficients in strangeness-neutral QCD matter at finite baryon density. The emergence of temperature-band crossings, temperature-hierarchy reversals, and correlated patterns among different coefficients reflects a nontrivial interplay between thermal and magnetic effects. Further studies with finer temperature resolution and extensions to higher orders in $\muB$ will be important for fully characterizing the QCD equation of state in strong magnetic backgrounds and for connecting lattice QCD predictions more directly to astrophysical and heavy-ion collision environments.

\section*{Acknowledgments}
This work is supported partly by the National Natural Science Foundation of China under Grants No. 12293064, No. 12293060, and No. 12325508, as well as the National Key Research and Development Program of China under Contract No. 2022YFA1604900. The numerical simulations have been performed on the GPU cluster in the Nuclear Science Computing Center at Central China Normal University ($\mathrm{NSC}^{3}$) and Wuhan Supercomputing Center.

\bibliographystyle{JHEP.bst}
\bibliography{refs.bib}

\appendix

\section{Lattice data and spline fits in $T$-$eB$ plane}
\label{app:data}

\begin{figure*}[!htbp]
\centering

\includegraphics[width=0.32\textwidth]{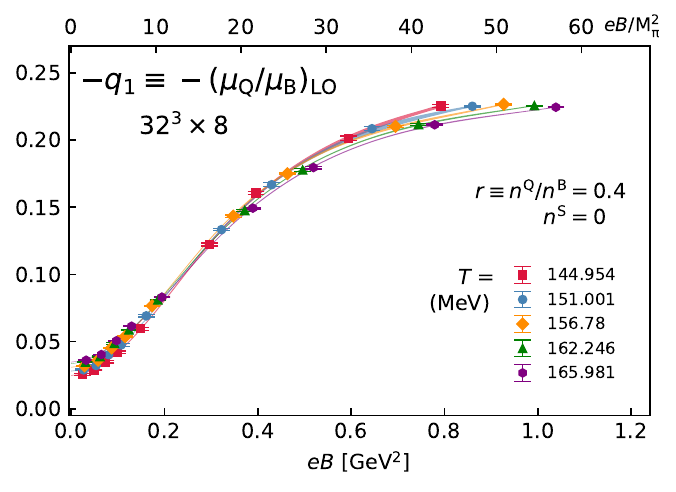}
\includegraphics[width=0.32\textwidth]{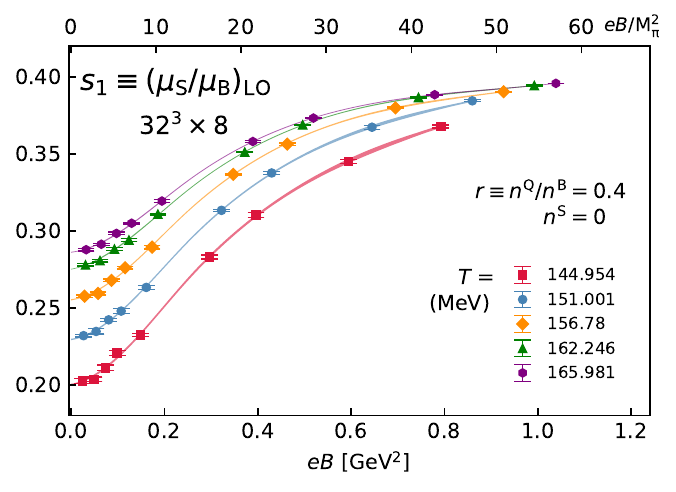}
\includegraphics[width=0.32\textwidth]{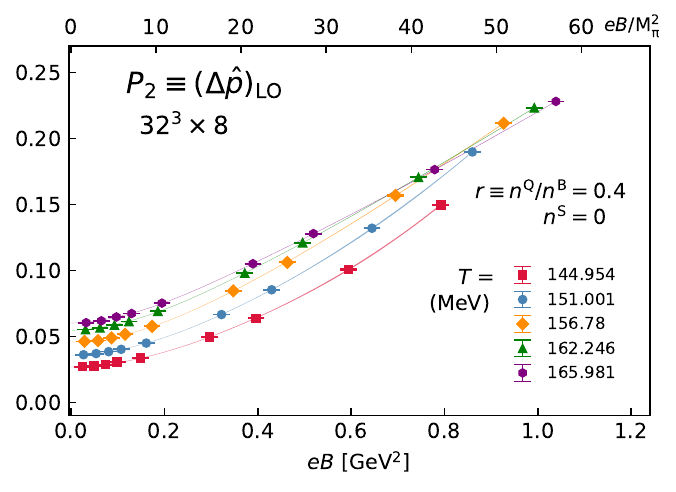}

\includegraphics[width=0.32\textwidth]{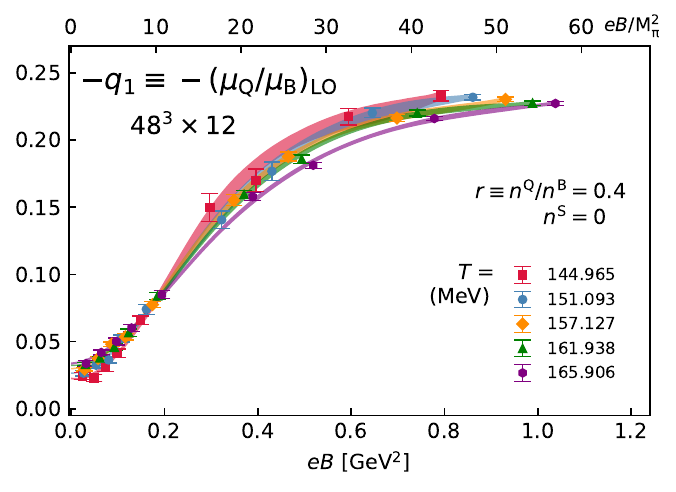}
\includegraphics[width=0.32\textwidth]{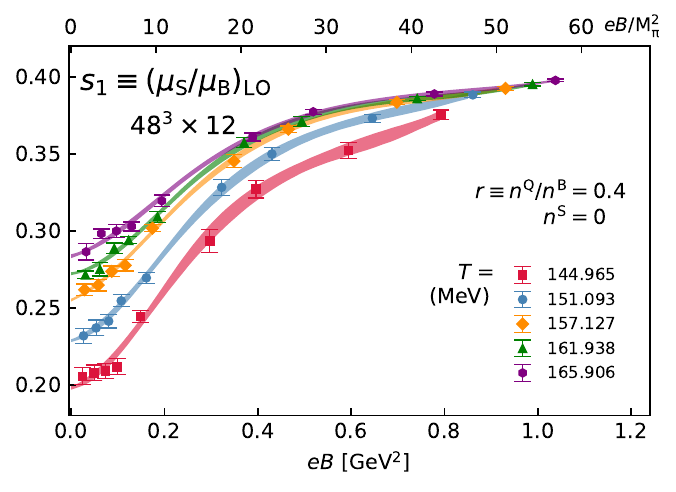}
\includegraphics[width=0.32\textwidth]{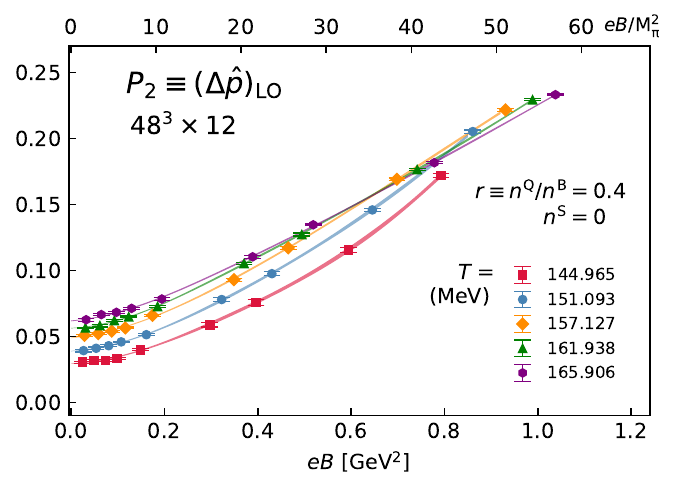}

\caption{Leading-order Taylor expansion coefficients  $-q_1\equiv -\left(\mu_{\rm Q}/\mu_{\rm B}\right)_{\rm LO}$ (left panel),  $s_1\equiv \left(\mu_{\rm S}/\mu_{\rm B}\right)_{\rm LO}$ (middle panel), and $P_2\equiv \left(\Delta \hat{p} \right)_{\rm LO}$ (right panel) obtained from lattice QCD simulations on lattices with temporal extents $N_{\tau}=8$ (top panel) and $N_\tau=12$ (bottom panel) spanning $T$-$eB$ parameter space. We imposed strangeness neutrality $\hat{n}^{\rm S}=0$ with slight isospin asymmetry $r\equiv n^\tQ/n^\tB =0.4$. The colored bands represent two-dimensional spline interpolations of lattice data points (filled symbols) at different temperatures.}   
\label{fig:vseB/lqcd_data_q1s1p2_nt812}
\end{figure*}

In this section, we present lattice QCD data results for leading-order Taylor expansion coefficients: $-q_1\equiv -\left(\mu_{\rm Q}/\mu_{\rm B}\right)_{\rm LO}$ (left panel),  $s_1\equiv \left(\mu_{\rm S}/\mu_{\rm B}\right)_{\rm LO}$ (middle panel), and $P_2\equiv \left(\Delta \hat{p} \right)_{\rm LO}$ in~\autoref{fig:vseB/lqcd_data_q1s1p2_nt812}. The top and bottom panels show results for $N_\tau=8$ and $N_\tau=12$ lattices, respectively, spanning the full $T$-$eB$ parameter space. These results correspond to a system with imposed strangeness neutrality ($\hat{n}^{\rm S}=0$), with slight isospin asymmetry ($r\equiv n^\tQ/n^\tB =0.4$). Distinct colored data points represent lattice QCD simulation results at fixed temperatures and various magnetic field strengths corresponding to the discrete magnetic flux values specified in Sec.~\ref{sec:setup}. Following the approach in Refs.~\cite{Bali:2011qj,Ding:2023bft}, we perform two-dimensional B-spline fits to the lattice data, shown as colored bands corresponding to the respective lattice data points. The average values and error bands are obtained using the Gaussian bootstrap method, with the median and the $68\%$ percentiles of the distribution, respectively.

The continuum estimates used in our work are obtained through linear extrapolations in $1/N_\tau^2$ using interpolated data at $ N_\tau = 8 $ and $ N_\tau = 12 $. For details, see Refs. \cite{Ding:2023bft, Ding:2025jfz}, where we demonstrated the consistency of these continuum estimates by including a supplementary $ N_\tau = 16 $ data point and performing an independent continuum extrapolation with $ N_\tau = 8 $, $ N_\tau = 12 $, and $ N_\tau = 16 $ results under the same $1/N_\tau^2 $ ansatz. Given the agreement within uncertainties, we adopt the continuum estimates from $ N_\tau = 8 $ and $12$ for our analyses, as they provide a statistically robust foundation while minimizing computational overhead.

\section{Parametrization of $q_1(T,eB)$, $s_1(T,eB)$, and $P_2(T,eB)$}
\label{app:param_eos}
In this section, we provide an analytical parameterization of the leading-order Taylor expansion coefficients—$q_1(T,eB)$, $s_1(T,eB)$, and $P_2(T,eB)$—obtained from our continuum-estimated $(2+1)$-flavor lattice QCD results in the presence of strong magnetic fields and nonzero baryon chemical potential. These parameterizations can be utilized for phenomenological applications and hydrodynamic modeling of strongly interacting matter.

For bulk thermodynamic Taylor expansion coefficient, $\mathcal{O}(T,eB)$ across two-dimensional $T$-$eB$ parameter space, we consider two distinct rational polynomial ansatze: $T$-parametrization at fixed magnetic strength $eB$, and $eB$-parametrization at fixed temperature $T$. The functional forms are similar in spirit to those used in the HotQCD EoS parameterizations at $eB=0$~\cite{HotQCD:2014kol,Bazavov:2017dus}. We find that fourth-order rational polynomials, in both parameterizations, accurately reproduce the lattice continuum-estimated lattice QCD results for the bulk thermodynamic Taylor expansion coefficients discussed in the previous sections.

$T$-parametrization: We employ the following fourth-order rational polynomial parametrization of $\mathcal{O}(T)$ with respect to temperature $T$, at various fixed magnetic field strengths, $eB \in \{0.0,~0.1,~0.2,~0.4,~0.6,0.79\}~{\rm GeV}^2$:
\beq
\label{eq:T-parametrization}
\mathcal{O}(T) \equiv \frac{\displaystyle \sum_{k=0}^{4}a_k \overline{T}^k}{1+ \displaystyle \sum_{k=1}^{4}b_k \overline{T}^k},
\eeq
where $\overline{T} = 1- \left(\frac{T_{pc}(eB=0)}{T} \right)$, and the QCD pseudo-critical temperature in the absence of magnetic fields, $T_{pc}(eB=0) = 155.029~{\rm MeV}$, is determined from the peak of the chiral susceptibility~\cite{Ding:2025jfz}. 

$eB$-parametrization: Following a similar procedure as above, we consider the parametrization of $\mathcal{O}(eB)$ with respect to magnetic field strength $eB$, evaluated at several fixed temperatures, $T \in \{145,~155,165\}~{\rm MeV}$:
\beq
\label{eq:eB-parametrization}
\mathcal{O}(eB) \equiv \frac{\displaystyle \sum_{k=0}^{4}a_k \overline{eB}^k}{1+ \displaystyle \sum_{k=1}^{4}b_k \overline{eB}^k},
\eeq
where $\overline{eB} = eB / eB_{\rm max}$, with $eB_{\rm max} = 0.8~{\rm GeV}^2$ corresponding to the largest magnetic field strength explored in our lattice continuum estimates.

\autoref{tab:parameterization_vsT} and \autoref{tab:parameterization_vseB} summarize the rational polynomial fit parameters $a_k$ and $b_k$ in the $T$- and $eB$-parametrizations, respectively, for the leading-order coefficients $\mathcal{O} \in \{ P_2,\,-q_1,\,s_1 \}$, evaluated for a strangeness-neutral system ($\hat{n}^{\rm S}=0$) with slight isospin asymmetry ($r \equiv n^{\rm Q}/n^{\rm B} = 0.4$).

\begin{table*}[!htbp]
    \centering
    \begin{tabular}{c|c|c|c|c|c|c|c|c|c|c}
        \hline \hline
        $\mathcal{O}(T)~~$ & $eB~({\rm GeV})^2$ &  
        $a_0$ & $a_1$   & $a_2$ & $a_3$& $a_4$ & 
        $b_1$ & $b_2$ & $b_3$ & $b_4$\\
        \hline 
        ${P_{2}}$ & 0.000 & 0.048 & -1.069 & 6.202 & 35.039 & 222.673 & -27.642 & 276.550 & -669.329 & 5464.583 \\
        & 0.100 & 0.056 & -0.089 & 0.922 & 60.324 & 180.559 & -6.288 & 71.458 & 922.847 & -5531.665 \\  
        & 0.200 & 0.071 & -0.484 & 9.745 & 122.980 & 601.250 & -10.480 & 193.319 & 1037.199 & 4460.654 \\
        & 0.400 & 0.109 & -3.011 & 36.507 & -80.767 & 95.526 & -30.090 & 429.869 & -2292.888 & 9648.112 \\
        & 0.600 & 0.153 & -2.453 & 40.185 & 154.508 & 3703.477 & -17.492 & 320.754 & 278.842 & 26625.567 \\
        & 0.790 & 0.199 & -1.761 & 12.634 & 156.305 & -1590.194 & -9.085 & 85.862 & 757.941 & -10296.849 \\        
        \hline
        $-{q_{1}}$ & 0.000 & 0.028 & -1.326 & 17.617 & 5.277 & -214.548 & -51.405 & 816.572 & -2759.911 & 3176.216 \\
        & 0.100 & 0.048 & -0.579 & -5.271 & 114.758 & 265.050 & -13.043 & -53.361 & 2330.893 & -4558.519 \\
        & 0.200 & 0.094 & -1.048 & -0.609 & 92.766 & 2978.945 & -11.646 & 17.402 & 1398.781 & 26078.533 \\
        & 0.400 & 0.179 & 1.778 & -13.785 & -147.964 & -814.930 & 10.124 & -78.749 & -618.848 & -1030.375 \\
        & 0.600 & 0.218 & -4.740 & 43.049 & 163.540 & -2797.055 & -20.732 & 166.262 & 761.575 & -7472.271 \\
        & 0.790 & 0.229 & -4.259 & 27.417 & 267.410 & -1812.098 & -17.502 & 89.661 & 1128.404 & -3044.080 \\
        \hline
        ${s_{1}}$ & 0.000 & 0.245 & -12.088 & 178.430 & 425.851 & -170.074 & -51.926 & 869.821 & -798.112 & 4609.446 \\
        & 0.100 & 0.274 & -2.016 & 9.695 & 355.918 & -80.230 & -9.823 & 90.379 & 1220.402 & -8084.585 \\
        & 0.200 & 0.309 & -1.620 & 15.627 & 480.668 & -2845.841 & -6.951 & 89.365 & 1602.061 & -16058.931 \\
        & 0.400 & 0.361 & 0.402 & 3.253 & 311.656 & -4396.561 & 0.434 & 20.229 & 933.628 & -14232.715 \\
        & 0.600 & 0.381 & -3.337 & -12.472 & 433.617 & 2381.702 & -9.138 & -21.506 & 1103.913 & 5233.861 \\
        & 0.790 & 0.388 & -4.579 & -3.625 & 1105.541 & -7239.611 & -11.957 & -2.075 & 2860.390 & -20174.486 \\
        \hline \hline	
    \end{tabular}
    \caption{$T$-parametrization: Fourth-order rational polynomial fit parameters from~\autoref{eq:T-parametrization} for the leading-order coefficients $P_2$, $-q_1$, and $s_1$ as functions of temperature. Fits are performed at various fixed magnetic strengths for a strangeness-neutral system ($\hat{n}^{\rm S}=0$) with slight isospin asymmetry ($r \equiv n^{\rm Q}/n^{\rm B} = 0.4$).}
    \label{tab:parameterization_vsT}
\end{table*}

\begin{table*}[!htbp]
    \centering
    \begin{tabular}{c|c|c|c|c|c|c|c|c|c|c}
        \hline \hline
        $\mathcal{O}(eB)~~$ & $T~({\rm MeV})$ &  
        $a_0$ & $a_1$   & $a_2$ & $a_3$& $a_4$ & 
        $b_1$ & $b_2$ & $b_3$ & $b_4$\\
        \hline 
        & 145 & 0.032 & 0.002 & 0.259 & -0.347 & 1.420 & -0.416 & -2.727 & 15.662 & -6.432 \\
        ${P_{2}}$ & 155 & 0.048 & -0.087 & 0.249 & -0.513 & 1.252 & -2.742 & 3.069 & 2.246 & 1.145 \\
        & 165 & 0.062 & 0.040 & 0.293 & 0.754 & 3.394 & 0.432 & -2.224 & 27.440 & -2.760 \\
         \hline
        & 145 & 0.021 & -0.024 & 2.250 & 2.328 & -3.325 & 4.342 & -13.446 & 41.901 & -28.628 \\
        $-{q_{1}}$ & 155 & 0.028 & 0.747 & 1.254 & 29.129 & -25.630 & 27.020 & -46.740 & 157.198 & -114.352 \\  & 165 & 0.033 & -0.084 & 1.208 & -2.453 & 1.435 & -2.803 & 7.515 & -11.236 & 6.142 \\        
        \hline
        & 145 & 0.196 & 2.420 & -2.655 & 23.319 & -17.095 & 11.916 & -26.735 & 88.776 & -58.796 \\
        ${s_{1}}$ & 155 & 0.245 & 0.039 & 1.034 & 0.856 & -1.939 & -0.530 & 2.653 & 2.717 & -5.232 \\
        & 165 & 0.279 & -0.950 & 3.065 & -5.697 & 3.644 & -3.591 & 9.581 & -15.709 & 9.598 \\        
        \hline
        \hline \hline	
    \end{tabular}
    \caption{$eB$-parametrization: Fourth-order rational polynomial fit parameters from~\autoref{eq:eB-parametrization} for the leading-order coefficients $P_2$, $-q_1$, and $s_1$ as functions of magnetic field strength. Fits are performed at several fixed temperatures for a strangeness-neutral system ($\hat{n}^{\rm S}=0$) with slight isospin asymmetry ($r \equiv n^{\rm Q}/n^{\rm B} = 0.4$).  }
    \label{tab:parameterization_vseB}
\end{table*}

\section{Parametrization of Second-Order Fluctuations and Correlations in Ref.~\cite{Ding:2025jfz}}
\label{app:param_flu}

In this section, we present an analytical parametrization for the continuum-estimated lattice QCD results for second-order conserved charge fluctuations in $(2+1)$-flavor QCD under a strong magnetic field, as reported in Ref.~\cite{Ding:2025jfz}. We adopt the same parametrization scheme as detailed in Appendix~\ref{app:param_eos}.

\autoref{tab:fit_params_T_flu} and~\autoref{tab:fit_params_eB_flu} summarize the rational polynomial fit parameters $a_k$ and $b_k$ for the $T$- and $eB$-dependent parameterizations, respectively, of the second-order fluctuations and correlations $\mathcal{O} \in \{ \chi_{2}^{\rm B},\, \chi_{2}^{\rm Q},\, \chi_{2}^{\rm S},\, \chi_{11}^{\rm BQ},\, \chi_{11}^{\rm BS},\, \chi_{11}^{\rm QS} \}$.

\begin{table*}[h]
\centering
\begin{tabular}{l|c|c|c|c|c|c|c|c|c|c}
\hline
\hline
$\mathcal{O}(T)$ & $eB~(\mathrm{GeV}^2)$ & $a_0$ & $a_1$ & $a_2$ & $a_3$ & $a_4$ & $b_1$ & $b_2$ & $b_3$ & $b_4$ \\
\hline
\quad                        & 0.10 & 0.136 & -1.862 & 0.732 & 177.324 & 1299.125 & -19.119 & 125.725 & 670.729 & 1737.377 \\
\quad                        & 0.30 & 0.242 & -5.841 & 14.022 & -381.888 & 742.655 & -27.820 & 174.518 & -2379.443 & 6916.062 \\
$\chi_{{ 2 }}^{{ \rm B }}$   & 0.50 & 0.392 & -13.599 & 146.454 & -170.289 & -1660.255 & -36.743 & 467.244 & -1954.232 & 4128.417 \\
\quad                        & 0.70 & 0.564 & -17.970 & 183.193 & 88.437 & -1907.504 & -32.494 & 364.856 & -644.909 & 2169.096 \\
\quad                        & 0.79 & 0.643 & -13.609 & 126.469 & 486.566 & 2131.769 & -21.209 & 212.256 & 479.184 & 5843.578 \\
\hline
\quad                        & 0.10 & 0.428 & 3.073 & 78.465 & 795.559 & 186.631 & 4.515 & 194.216 & 1530.326 & -6256.305 \\
\quad                        & 0.30 & 0.584 & -10.398 & 29.606 & -501.009 & 14789.493 & -20.984 & 133.875 & -1442.260 & 24663.500 \\
$\chi_{{ 2 }}^{{ \rm Q }}$   & 0.50 & 0.872 & 32.025 & -204.833 & -3064.394 & -25282.990 & 34.866 & -291.383 & -2619.442 & -24790.769 \\
\quad                        & 0.70 & 1.263 & -16.978 & -197.124 & 1644.940 & 26848.014 & -13.721 & -146.727 & 1178.162 & 22049.333 \\
\quad                        & 0.79 & 1.463 & 15.365 & 113.017 & -2437.434 & -29683.750 & 10.879 & 85.110 & -1761.097 & -19587.816 \\
\hline
\quad                        & 0.10 & 0.296 & 0.258 & 48.761 & 578.107 & 1783.140 & -4.050 & 193.292 & 1075.117 & -3033.742 \\
\quad                        & 0.30 & 0.380 & -1.212 & -15.185 & 456.541 & 2028.503 & -7.796 & -16.636 & 1314.060 & 1242.161 \\
$\chi_{{ 2 }}^{{ \rm S }}$   & 0.50 & 0.532 & -0.372 & -31.727 & 658.423 & 4940.868 & -4.058 & -50.779 & 1282.936 & 6570.386 \\
\quad                        & 0.70 & 0.739 & -2.296 & -21.187 & 2322.955 & 14837.072 & -5.268 & -7.869 & 2960.267 & 13026.339 \\
\quad                        & 0.79 & 0.845 & 0.523 & 21.144 & 1722.285 & 9983.241 & -1.122 & 38.713 & 1801.459 & 8145.362 \\
\hline
\quad                        & 0.10 & 0.038 & -0.332 & 0.130 & 74.551 & 404.005 & -11.943 & 107.269 & 1693.778 & -3001.360 \\
\quad                        & 0.30 & 0.118 & -2.092 & -0.737 & 19.253 & 818.063 & -19.873 & 42.038 & 290.302 & 1462.219 \\
$\chi_{{ 11 }}^{{ \rm BQ }}$ & 0.50 & 0.235 & -5.208 & 96.605 & -30.838 & -10997.235 & -22.417 & 412.710 & -493.881 & -38320.972 \\
\quad                        & 0.70 & 0.366 & -3.188 & 52.024 & -97.757 & -6655.676 & -7.593 & 130.114 & -496.089 & -11699.542 \\
\quad                        & 0.79 & 0.426 & -4.104 & 70.383 & 78.161 & -7567.477 & -8.123 & 151.435 & 71.763 & -12602.321 \\
\hline
\quad                        & 0.10 & -0.075 & 0.714 & -0.639 & -96.243 & -985.503 & -17.041 & 150.793 & 421.325 & 2078.510 \\
\quad                        & 0.30 & -0.113 & 0.466 & -10.240 & -391.231 & -1723.200 & -10.109 & 159.226 & 2765.310 & -3219.697 \\
$\chi_{{ 11 }}^{{ \rm BS }}$ & 0.50 & -0.169 & -0.748 & -0.645 & -534.036 & -2595.324 & 0.293 & 17.646 & 3155.351 & 449.354 \\
\quad                        & 0.70 & -0.241 & 3.227 & 5.404 & -528.608 & -2818.454 & -16.040 & 37.041 & 1816.625 & 7104.233 \\
\quad                        & 0.79 & -0.277 & 3.885 & -14.379 & -260.913 & 83.904 & -16.116 & 100.879 & 436.195 & 262.391 \\
\hline
\quad                        & 0.10 & 0.107 & -0.099 & 9.106 & 11.423 & 52.545 & -5.291 & 96.094 & -262.324 & 1772.356 \\
\quad                        & 0.30 & 0.120 & -2.351 & 21.508 & -56.850 & -2215.755 & -24.549 & 282.296 & -1504.506 & -9410.380 \\
$\chi_{{ 11 }}^{{ \rm QS }}$ & 0.50 & 0.148 & 0.784 & 32.517 & 250.078 & -372.909 & 1.815 & 206.696 & 646.469 & -2817.426 \\
\quad                        & 0.70 & 0.197 & 2.016 & -6.598 & 188.853 & 2044.960 & 8.283 & -35.267 & 799.821 & 3428.472 \\
\quad                        & 0.79 & 0.226 & -2.140 & 12.329 & 105.710 & -1847.307 & -11.042 & 91.948 & -183.607 & -5408.080 \\
\hline
\hline
\end{tabular}
\caption{$T$-parametrization: Fourth-order rational polynomial fit parameters from~\autoref{eq:T-parametrization} for the second-order fluctuations and correlations $\{ \chi_{2}^{\rm B},\, \chi_{2}^{\rm Q},\, \chi_{2}^{\rm S},\, \chi_{11}^{\rm BQ},\, \chi_{11}^{\rm BS},\, \chi_{11}^{\rm QS} \}$  as functions of temperature.}
\label{tab:fit_params_T_flu}
\end{table*}

\begin{table*}
\centering
\begin{tabular}{l|c|c|c|c|c|c|c|c|c|c}
\hline
\hline
$\mathcal{O}(eB)$ & $T~(\mathrm{MeV})$ & $a_0$ & $a_1$ & $a_2$ & $a_3$ & $a_4$ & $b_1$ & $b_2$ & $b_3$ & $b_4$ \\
\hline
\quad                        & 145 & 0.071 & -0.088 & 1.142 & -2.311 & 4.096 & -1.239 & -0.287 & 8.130 & -2.861 \\
$\chi_{{ 2 }}^{{ \rm B }}$   & 155 & 0.113 & -0.176 & 1.071 & -2.275 & 3.802 & -2.220 & 2.659 & 1.254 & 1.194 \\
\quad                        & 165 & 0.155 & -0.290 & 1.450 & -3.156 & 8.775 & -1.890 & 0.332 & 10.632 & 1.010 \\
\hline
\quad                        & 145 & 0.340 & -1.055 & 3.513 & -3.970 & 7.042 & -2.717 & 5.836 & -0.714 & 0.858 \\
$\chi_{{ 2 }}^{{ \rm Q }}$   & 155 & 0.409 & -0.295 & 3.386 & -1.082 & 6.799 & -0.650 & 4.637 & 0.781 & 0.435 \\
\quad                        & 165 & 0.464 & -0.944 & 4.341 & -5.060 & 11.047 & -1.800 & 4.091 & 1.897 & 1.666 \\
\hline
\quad                        & 145 & 0.194 & -0.396 & 1.435 & -1.134 & 3.096 & -1.991 & 3.727 & 2.912 & -0.934 \\
$\chi_{{ 2 }}^{{ \rm S }}$   & 155 & 0.276 & -0.662 & 1.310 & -1.048 & 1.479 & -2.763 & 4.236 & -1.589 & 0.698 \\
\quad                        & 165 & 0.371 & -0.828 & 1.910 & -1.743 & 6.066 & -2.186 & 1.725 & 5.951 & -0.343 \\
\hline
\quad                        & 145 & 0.017 & -0.065 & 1.231 & -2.242 & 3.301 & -0.584 & -1.085 & 9.177 & -3.397 \\
$\chi_{{ 11 }}^{{ \rm BQ }}$ & 155 & 0.024 & -0.053 & 0.847 & -2.698 & 3.499 & -2.740 & 2.484 & 1.338 & 1.662 \\
\quad                        & 165 & 0.028 & -0.065 & 0.902 & -2.472 & 2.579 & -2.360 & 2.097 & -0.395 & 2.144 \\
\hline
\quad                        & 145 & -0.036 & 0.043 & -0.332 & 0.456 & -1.150 & -1.471 & 1.486 & 6.146 & -2.487 \\
$\chi_{{ 11 }}^{{ \rm BS }}$ & 155 & -0.065 & 0.123 & -0.217 & 0.652 & -0.990 & -2.629 & 1.189 & 2.442 & -0.232 \\
\quad                        & 165 & -0.099 & 0.054 & -0.660 & -0.293 & -3.724 & -0.392 & -0.118 & 18.409 & -3.609 \\
\hline
\quad                        & 145 & 0.079 & -0.088 & 0.118 & 0.116 & -0.159 & -1.368 & 3.604 & -4.149 & 1.303 \\
$\chi_{{ 11 }}^{{ \rm QS }}$ & 155 & 0.106 & 4.482 & 4.997 & 0.780 & 5.700 & 42.444 & 46.993 & -30.224 & 9.938 \\
\quad                        & 165 & 0.136 & -0.500 & 0.720 & -0.398 & 0.762 & -3.654 & 4.067 & 1.612 & -0.201 \\
\hline
\hline
\end{tabular}
\caption{$eB$-parametrization: Fourth-order rational polynomial fit parameters from~\autoref{eq:eB-parametrization} for the second-order fluctuations and correlations $\{ \chi_{2}^{\rm B},\, \chi_{2}^{\rm Q},\, \chi_{2}^{\rm S},\, \chi_{11}^{\rm BQ},\, \chi_{11}^{\rm BS},\, \chi_{11}^{\rm QS} \}$  as functions of magnetic field strength.}
\label{tab:fit_params_eB_flu}
\end{table*}

\end{document}